\documentclass[graybox,envcountchap]{svmono}

\usepackage{type1cm}         

\usepackage{makeidx}         
\usepackage{graphicx}        
\usepackage{multicol}        
\usepackage[bottom]{footmisc}

\usepackage{newtxtext}       %
\usepackage[varvw]{newtxmath}       

\usepackage{amsmath}
\usepackage{comment}
\usepackage{adjustbox}

\newcommand{\ali}[1]{\begin{align} #1 \end{align}}
\newcommand{\p}{\partial}
\newcommand{\ra}{\rightarrow}

\newcommand{\vev}[1]{\langle #1 \rangle} 
\newcommand{\mn}{{\mu\nu}}

\DeclareMathOperator\tr{\text{tr}}

\definecolor{glueblue}{rgb}{0, 0.44, 1} 
\definecolor{taugreen}{rgb}{0, 0.345, 0.149} 
\definecolor{islandblue}{rgb}{0.267, 0.549, 0.796}  
\definecolor{radiationred}{rgb}{0.917, 0, 0.455}  
\definecolor{contourpink}{rgb}{0.94, 0.43, 0.67} 
\definecolor{tSblue}{rgb}{0, 0.33, 0.65}

\usepackage[hidelinks]{hyperref} %

\makeindex             

\begin{document}

\author{Nele Callebaut} 
\title{Entanglement in conformal field theory and holography}
\maketitle 

\abstract{In these notes we give a pedagogical account of the replica trick derivation of CFT entanglement and its holographic counterpart, i.e.~the Lewkowycz Maldacena derivation of the Ryu-Takayanagi formula. The application to an `island set-up' for the calculation of black hole radiation entropy is briefly discussed. Further topics focused on are the relation to thermal entropy, thermofield double constructions and statements about the emergence of gravity from entanglement through reinterpretations of gravitational first laws.}



\tableofcontents

\mainmatter
\chapter{CFT entanglement}	\label{chapterCFT}

\abstract*{Entanglement entropy in a conformal field theory can be defined as the von Neumann entropy of the reduced density matrix. We review in this chapter the replica trick derivation of CFT entanglement, following the work of Cardy and Calabrese in section \ref{sectionreplicaCFT} and of Holzhey, Larsen and Wilczek in section \ref{sectionThermal}. This follows an introductory section \ref{sectintro} covering the pictorial notation of wave-functionals and density matrices in the path integral formalism, which is also heavily used in section \ref{sectTFD} on the thermofield double construction. }

Entanglement entropy in a conformal field theory (CFT) can be defined as the von Neumann entropy of the reduced density matrix. We review in this chapter the replica trick derivation of CFT entanglement, following the work of Cardy and Calabrese \cite{Calabrese:2004eu,Calabrese:2009qy,Cardy:2008jc} in section \ref{sectionreplicaCFT} and of Holzhey, Larsen and Wilczek \cite{Holzhey:1994we} in section \ref{sectionThermal}. This follows an introductory section \ref{sectintro} covering the pictorial notation of wave-functionals and density matrices in the path integral formalism, which is also heavily used in section \ref{sectTFD} on the thermofield double construction.

\section{Wave-functionals and density matrices} \label{sectintro}

We start with reviewing\footnote{For more details, see Appendix A of \cite{Polchinski:1998rq} and section 4 of \cite{Hartmanlectures}.} some basic pictorial notation for wave-functionals and density matrices that will be used throughout. 

In a quantum mechanical system with Hamiltonian $H$ and Euclidean action $I$, the transition amplitude from a position eigenstate at Euclidean time $t_E = -T$ to another position eigenstate at $t_E = 0$ can be written as a Euclidean path integral    
\ali{
	\langle q_f| e^{-H \, T}|q_i \rangle = \int_{q_i, -T}^{q_f,0} \mathcal D q e^{-I}.  
}
Inserting a complete set of energy eigenstates on the left hand side and taking the limit  
$T \ra \infty$ picks out the vacuum state contribution $|\psi\rangle$,   
such that the wavefunction $\psi(q_f)$ is given by path integral evolution from past Euclidean infinity (at a fixed and unimportant initial position $q_i$)    
\ali{
	\psi(q_f) = \int_{q_i, t_E = -\infty}^{q_f,t_E = 0} \mathcal Dq \, e^{-I} , 
}
where $q_f$-independent factors have been dropped. The ket $|\psi\rangle$ is the function  
\ali{
	|\psi\rangle = \int_{q_i, t_E = -\infty}^{\cdot,t_E = 0} \mathcal Dq \, e^{-I}   
}
that takes a position $q_f$ (at the dot $\cdot$) and gives back a complex number $\psi(q_f) = \langle q_f|\psi\rangle$.  
On slices of the path integral one recovers the Hilbert space of the theory. 

Similarly, in a quantum field theory with field content $\phi(\vec x,t)$, the (unnormalized) vacuum state is prepared by Euclidean evolution 
\ali{ 
	|\psi \rangle = \int_{\phi(t_E = -\infty) = \phi_i}^{\phi(t_E = 0) = \,  \cdot} \mathcal D\phi \, e^{-I} \quad = \quad  \adjincludegraphics[width=2cm,valign=c]{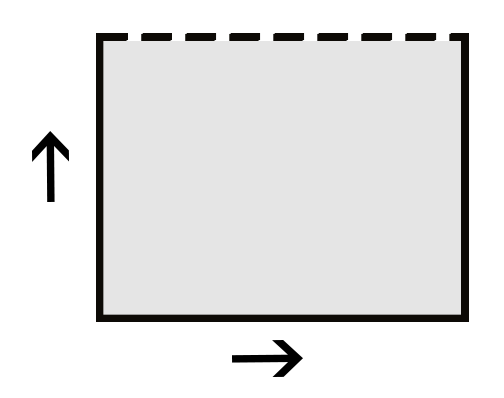}  \llap{
		\parbox[][-1cm][b]{2.1cm}{$t_E$}} \llap{
		\parbox[][1.5cm][b]{0.5cm}{$x$}},	\label{psiket}
}
where we have already for concreteness restricted the pictorial representation to field theories with one spatial direction $x$.  
The state can then further be evolved in Lorentzian time $t$, with the wave-functional   
\ali{
	\psi(\phi_f) = \int_{\phi(t_i) = \phi_i}^{\phi(t_f) = \phi_f} \mathcal D \phi \, e^{i \mathcal I}  = \int_{\phi(t_i) = \phi_i}^{\phi(t_f) = \phi_f} \mathcal D \phi \, e^{i \int_{t_i}^{t_f} L[\phi]}  
}
shown in \cite{Larsen:1994yt} to satisfy the Schr\"odinger evolution equation. Again, the ket for the vacuum state \eqref{psiket} is a functional  with open boundary condition (at the dot $\cdot$) where it can take on a given field configuration $\phi_f$ and give back 
\ali{
	\langle \phi_f | \psi \rangle = \int_{\phi(t_E = -\infty) = \phi_i}^{\phi(t_E = 0) = \phi_f} \mathcal D\phi \, e^{-I} = \adjincludegraphics[width=1.5cm,valign=c]{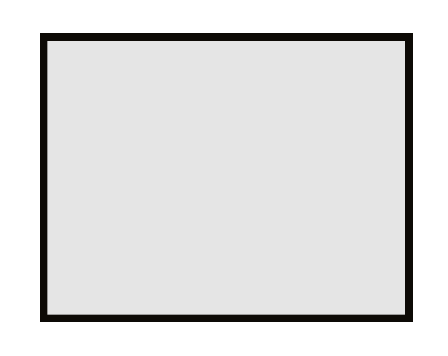}\llap{
		\parbox[][-1.1cm][b]{0.9cm}{$\phi_f$}}  ,  \label{psiphif}
}
obtained pictorially by gluing in the bra $\langle \phi_f |$. 

The corresponding bra $\langle \psi|$ and (unnormalized) density matrix $\rho = |\psi \rangle \langle \psi|$ for the pure vacuum state can then pictorially be presented as 
\ali{
	\langle \psi| = \adjincludegraphics[width=1.5cm,valign=c]{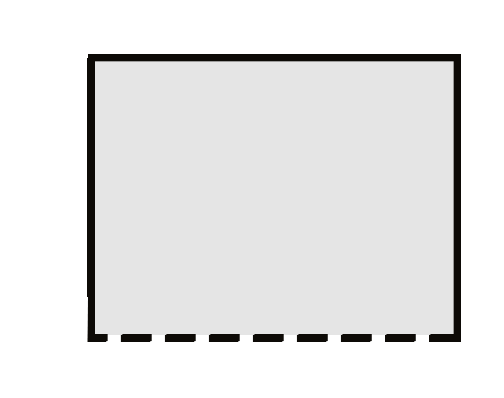}, \qquad \rho = \adjincludegraphics[width=1.5cm,valign=c]{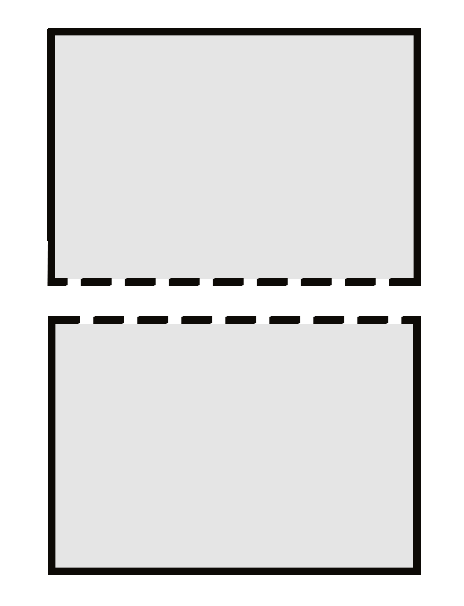} . \label{rhofig}
}
An upper dashed line can take on a bra, a lower dashed line a ket, and the matrix $\rho$ both, with matrix elements 
\ali{
	\langle \phi| \rho | \phi' \rangle \equiv (\rho)_{\phi \phi'} = \adjincludegraphics[width=1.5cm,valign=c]{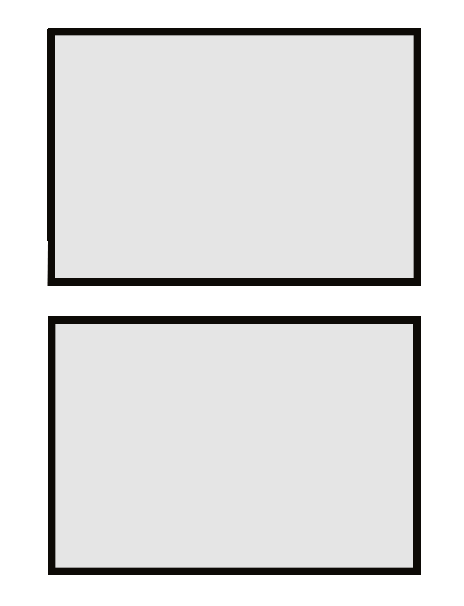}\llap{
		\parbox[][-.25cm][b]{0.85cm}{$\phi'$}} \llap{
		\parbox[][0.75cm][b]{0.85cm}{$\phi$}}  .  \label{rhophiphip}
}
Its trace $\tr \rho = \int \mathcal D \phi \langle \phi| \rho |\phi \rangle$, obtained by gluing along the previously dashed lines to identify and integrate out $\phi$,  
gives the partition function $Z$, 
\ali{
	Z = \tr \rho =  {\color{glueblue} \int \mathcal D \phi}   \adjincludegraphics[width=1.5cm,valign=c]{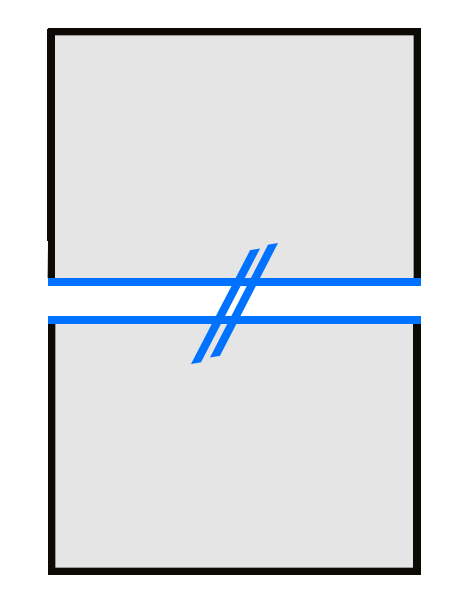} \llap{
		\parbox[][-0.25cm][b]{1.2cm}{${\color{glueblue}\phi}$}} \llap{
		\parbox[][0.75cm][b]{1.2cm}{${\color{glueblue}\phi}$}}  = \adjincludegraphics[width=1.5cm,valign=c]{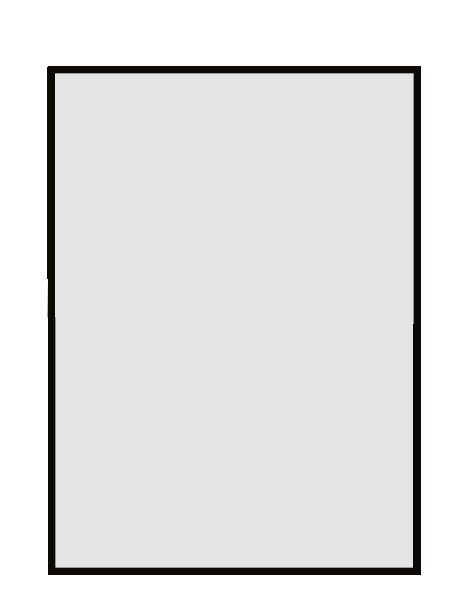} . \label{trrhoblue} 
}

In some figures of states \eqref{psiket} and density matrices \eqref{rhofig}  throughout in the text we will include field configurations in brackets to indicate how to read off the corresponding wave-functional \eqref{psiphif} and density matrix elements \eqref{rhophiphip}.

\section{CFT entanglement from replica trick} \label{sectionreplicaCFT}

For a given statistical ensemble described by $\rho$, where $\rho$ is the probability distribution classically or density matrix quantum mechanically, the von Neumann entropy is defined in terms of the normalized density matrix $\hat \rho = \rho/\tr \rho = \rho/Z$ as 
\ali{
	S = - \tr (\hat \rho \log \hat \rho).   
}
It provides a fundamental, observer-dependent measure for the indeterminacy or lack of resolution of the system, e.g.~$S = k_B \log \Omega(E)$ in the microcanonical ensemble, for an observer in a closed system, in which every microstate is equally probable $\hat \rho = 1/\Omega(E)$, or $S = (1 - \beta \p_\beta) \log Z(\beta)$ for an open system observer in the canonical ensemble with Boltzmann probability distribution $\hat \rho = e^{-\beta H}/Z(\beta)$.

The von Neumann entropy 
can be applied in the context of a conformal field theory to define the concept of `geometric entropy' or  entanglement entropy. 

The set-up we consider is a $(1+1)$-dimensional CFT with Euclidean path integral 
\ali{ 
	Z = \int \mathcal D \phi \, e^{-I[\phi]},  \qquad I[\phi] = \int_\mathbb{C} dx dt_E\mathcal L[\phi(x,t_E)],  
} 
prepared in a pure state $|\psi\rangle$ as pictured in \eqref{psiket}. 
The corresponding density matrix $\rho = |\psi\rangle \langle \psi|$ is  pictured in \eqref{rhofig}. We consider a constant time slice and geometrically bipartition the system, assuming the Hilbert space can be factorized, into a spatial region $A$ and its complement $\bar A$. 
An observer that only has access to region $A$ will measure a different density matrix, called the reduced density matrix 
\ali{
	\rho_A = \tr_{\bar A} \rho.  
}
It is obtained from $\rho$ by tracing out degrees of freedom in $\bar A$, and in general takes the form of a density matrix for a mixed state. 
The observer's lack of information about the full system can be quantified by the von Neumann entropy of the normalized reduced density matrix $\hat \rho_A = \rho_A/ Z$,  
\ali{
	S_A = - \tr (\hat \rho_A \log \hat \rho_A ) . \label{SAdef} 
}
This is by definition the \emph{geometric entropy} or \emph{entanglement entropy} associated with region $A$. 
It is a measure for the amount of missing information from the point of view of the observer in $A$, vanishing in the limit that the observer has access to the full system since the von Neumann entropy for a pure density matrix  is zero, and a measure for the amount of entanglement between degrees of freedom in $A$ and degrees of freedom in $\bar A$, vanishing when the pure state of the CFT is separable, $|\psi \rangle = |\psi_A \rangle|\psi_{\bar A} \rangle$ and $\rho_A= |\psi_A \rangle \langle \psi_A|$, and there is thus no such entanglement. 

From applying l'H\^{o}pital's rule, the definition for the entanglement entropy $S_A$ in \eqref{SAdef} can be rewritten as  
\ali{
	S_A &= (1 - n \p_n) \log Z(n)|_{n \ra 1},  \label{Sformula} \\
	Z(n) &= \tr \, (\rho_A^n) .  \label{eqZn} 
}
A positive integer $n$ factors of $\rho_A$ construct $Z(n)$, an $n$-fold replicated description of the system. Then $n$ needs to be analytically continued to non-integer values of $n$ to be able to take the derivative and limit in the definition \eqref{Sformula} of $S_A$. This is called the replica method.

\begin{figure}
	\centering \includegraphics[scale=0.15]{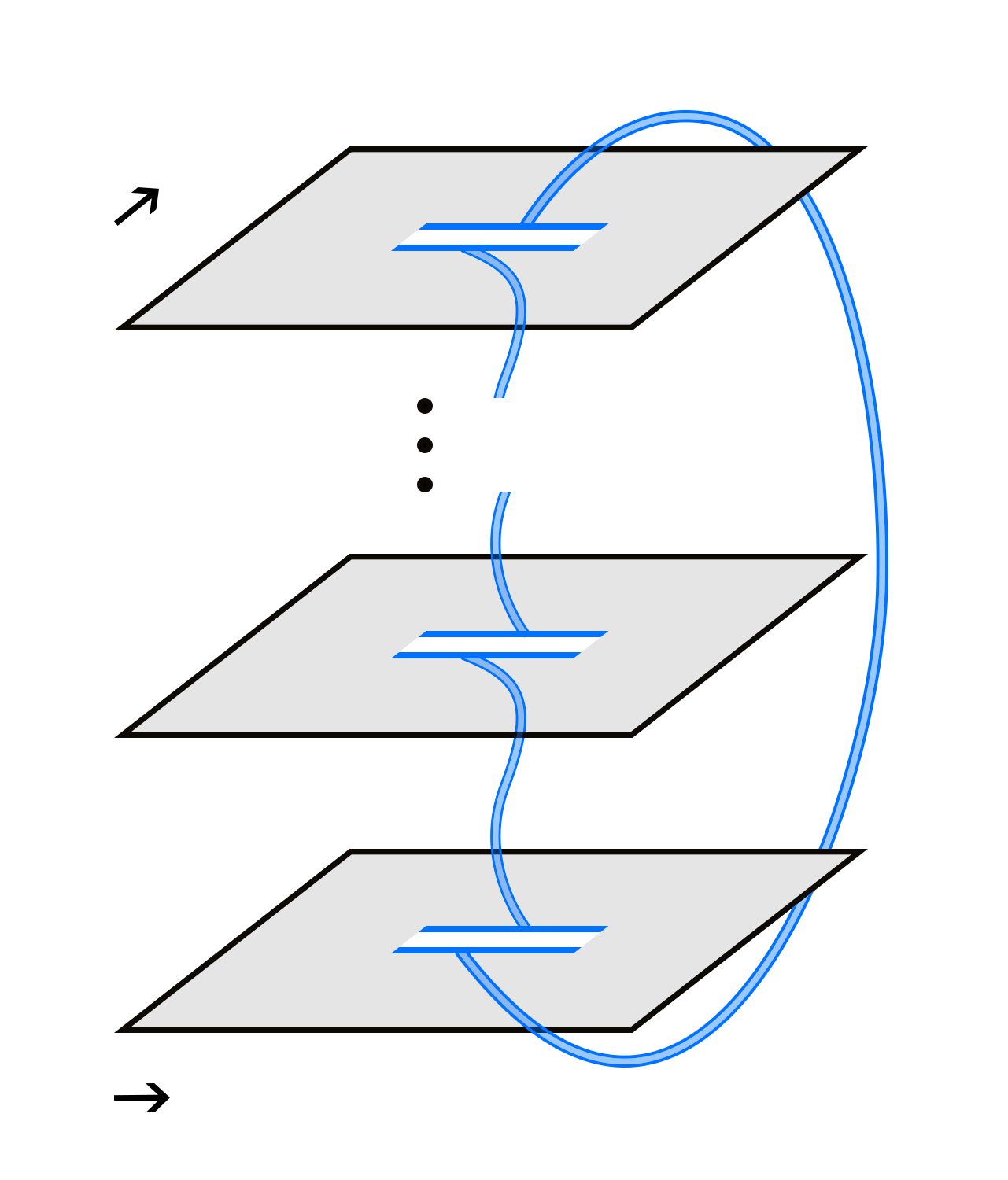} \llap{
		\parbox[][-10cm][b]{4.3cm}{$t_E$}} \llap{
		\parbox[][-0.5cm][b]{4.1cm}{$x$}}  \qquad \includegraphics[scale=0.15]{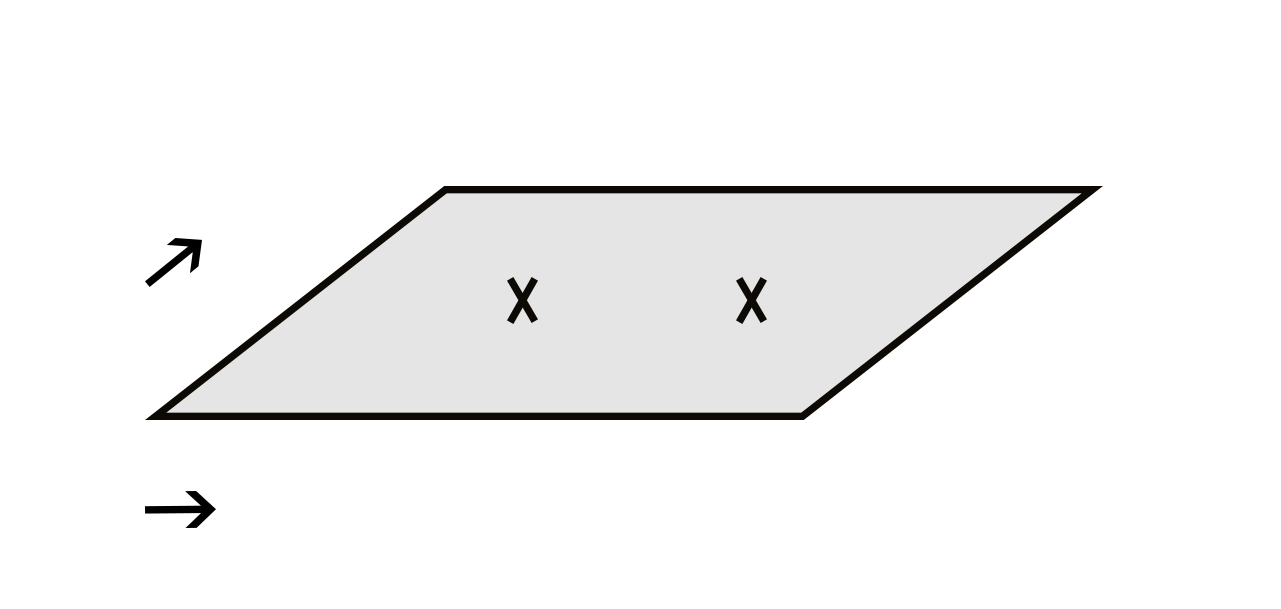} \llap{
		\parbox[][-2.9cm][b]{4.2cm}{$t_E$}} \llap{
		\parbox[][-0.5cm][b]{4.1cm}{$x$}}
	\caption{$Z(n) = \tr \rho_A^n$ from the WS perspective \eqref{WSZn} (left) and the TS perspective \eqref{TSZn} (right). It is the path integral of the theory $\mathcal L[\phi]$ on the $\mathbb Z_n$ symmetric replica manifold $\mathcal R_{n,A}$, or equivalently the path integral of the theory $\mathcal L^{(n)}[\phi_i]$ over the orbifold manifold $\mathbb C \equiv \mathcal R_{n,A}/\mathbb Z_n$ in the presence of twist fields. On the right, $Z(n) = \langle T_n(x_1) \tilde T_n(x_2)\rangle$. 
	} 
	\label{figreplica}
\end{figure}

Now consider region $A$ to be an interval $x = x_1..x_2$ at $t_E=0$, with corresponding reduced density matrix $\rho_A$ and its square $\rho_A^2$ given by 
\ali{
	\rho_A = \adjincludegraphics[width=2cm,valign=c]{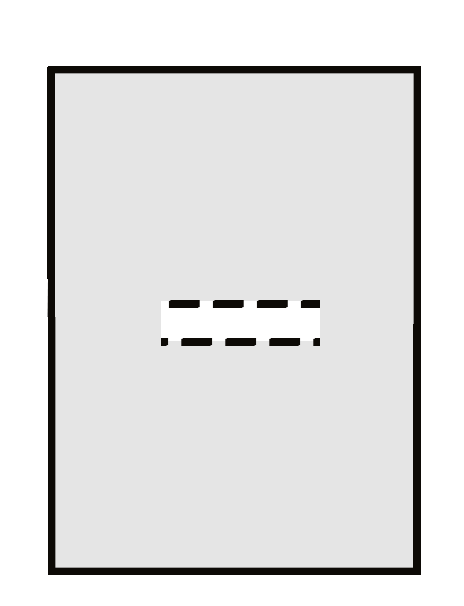}, \qquad \rho_A^2 = {\color{glueblue} \int \mathcal D \phi}   \adjincludegraphics[width=4cm,valign=c]{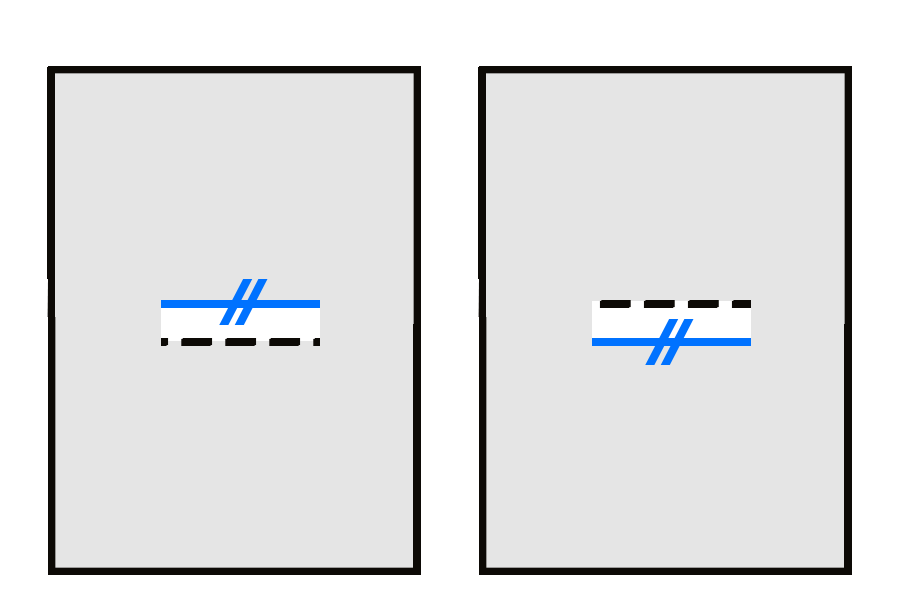} \llap{
		\parbox[][-0.4cm][b]{3cm}{${\color{glueblue}\phi}$}}  
	\llap{
		\parbox[][-0.65cm][t]{1.1cm}{${\color{glueblue}\phi}$}}. 
}    
To calculate the corresponding entanglement $S_A$ in \eqref{Sformula}, we need to construct $Z(n) = \tr \rho_A^n$ for integer $n>1$. 
We can think of the gluing condition (referring  
to the identification and integrating out of the field configuration) in the matrix multiplication $\rho_A^2$ in two ways: 1) as continuing the \emph{coordinates} $(t_E,x)$ to live on a connected manifold consisting of two copies of the complex plane glued along the region $A$ slit,  
or 2) as a condition on the \emph{field content}, connecting the fields of two separate copies of the theory $\mathcal L(\phi_1)$ and $\mathcal L(\phi_2)$ along $A$. 
The first is a `worldsheet' and the second a `target space' perspective, with $(t_E,x)$ running over $\mathcal R_{n,A}$ and $\mathbb{C}$ respectively. 
Constructing $\tr \rho_A^n$ in the first way  
gives rise to the replicated worldsheet  
$\mathcal R_{n,A}$, path integral integration over which gives $Z(n)$ 
\ali{
	\tr \rho_A^n = Z(n) &= \int [\mathcal D\phi]_{_{\mathcal R_{n,A}}} \, e^{-\int_{\mathcal R_{n,A}} dx dt_E \mathcal L[\phi(x,t_E)]} . \label{WSZn} 
}
$\mathcal R_{n,A}$ is called the replica manifold and is pictured in the left figure of Fig \ref{figreplica}. In the $Z(1)$ manifold a rotation over $2\pi$ will bring you back to the same location, but in the $Z(n)$ manifold it takes a rotation of $2\pi n$ around the boundary points $\p A$ of region $A$ to get back to the same location. That is, there are branch points and conical singularities at $\p A$, with a conical excess of $2\pi n - 2\pi$.

In the second perspective we write (here $i = 1 \cdots n$) 
\ali{
	\tr \rho_A^n = Z(n) &= \int [\mathcal D\phi_i]_{_{\mathbb{C},\, bc}} \, e^{-\int_\mathbb{C} dx dt_E\mathcal L^{(n)}[\phi_i(x,t_E)]} \label{TSZn}
}
with 	
\ali{ 
	\mathcal L^{(n)}[\phi_1, ..., \phi_n] &=  \mathcal L[\phi_1(x,t_E)] + \cdots + \mathcal L[\phi_n(x,t_E)] \\  
	bc &= \left\{ \begin{array}{l} \phi_1(t_E = 0^+, x \in A) = \phi_2(t_E=0^-, x\in A) \\  \phi_2(t_E = 0^+, x \in A) = \phi_3(t_E=0^-, x\in A) \\ \cdots \\ \phi_n(t_E = 0^+, x \in A) = \phi_1(t_E=0^-, x\in A) \end{array} \right. . 
}	
The boundary conditions $bc$ express a global symmetry of the theory $\mathcal L^{(n)}[\phi_1, ..., \phi_n]$ under exchange of the fields $\phi_i \ra \phi_{i+1}$,  the $\mathbb{Z}_n$ permutation symmetry. 
The conditions can be implicitly implemented by placing twist fields at $\p A$. These have the property that when circling a twist field $T_n$ resp. anti twist field $\tilde T_n$, a field $\phi_{i \text{ mod } n}$ is transformed into $\phi_{i+1 \text{ mod } n}$, resp. to $\phi_{i-1 \text{ mod } n}$. Then,    
\ali{
	Z(n) &= \int [\mathcal D\phi_i]_{_{\mathbb C}} \,\, T_n(x_1) \, \tilde  T_n(x_2)  \, e^{-\int_\mathbb{C} dx dt_E\mathcal L^{(n)}[\phi_i(x,t_E)]} \nonumber  \\
	&= \langle  T_n(x_1) \, \tilde T_n(x_2) \rangle_{\mathcal L^{(n)},\mathbb C}  \label{TSZnbis}
}
where we used a condensed notation for the locations of the twist fields $T_n(x=x_1,t_E=0)$ and $\tilde  T_n(x=x_2,t_E=0)$,   
and in the second line we rewrite the $Z(n)$ partition function as a 2-point function of twist fields. This interpretation of $Z(n)$ is pictured on the right of Fig \ref{figreplica}.

\subsection{Replica manifold} \label{subsectreplica}

We will proceed first with calculating $Z(n)$ in the replicated worldsheet point of view of Fig \ref{figreplica}a, following \cite{Cardy:2008jc}, and comment on the twist field correlator perspective later.

For a theory with stress tensor defined via $\delta I = \frac{1}{4\pi} \int T_\mn \delta g^\mn \sqrt{g}\, d^d x$, 
the partition function satisfies $\delta \log Z = -\frac{1}{4\pi} \int \vev{T_\mn} \delta g^\mn \sqrt{g}\, d^d x$. This allows us to write down 
the behavior of $Z(n)$ under a coordinate transformation $x^\mu \ra x'^\mu = x^\mu + \delta x^\mu$ (which induces a metric transformation $\p_\mu \delta x_\nu + \p_\nu \delta x_\mu$)
\ali{
	\delta \log Z(n) = - \frac{1}{2\pi} \int \vev{T^\mu_{\phantom{\mu}\nu}} \frac{\p \delta x^\nu}{\p x^\mu} d^2 x . \label{deltalogZn}
} 
This is for the theory $\mathcal L[\phi(x,t_E)]$ on the replica manifold $\mathcal R_{n,A}$ spanned by $x^\mu = (x,t_E)$, which is everywhere flat except at the branch points $\p A = (x_1,x_2)$. 

\begin{figure}
	\centering \includegraphics[scale=0.16]{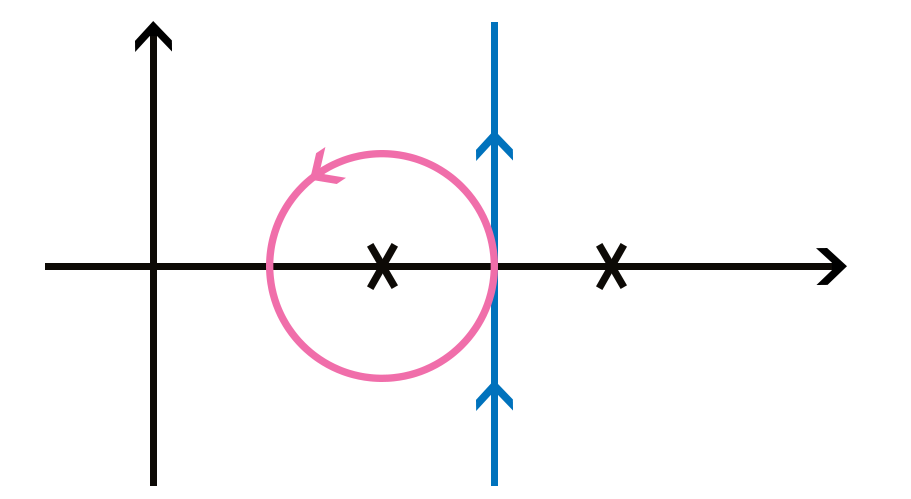} \llap{
		\parbox[][-3.6cm][b]{3.75cm}{$t_E$}} \llap{
		\parbox[][-1.5cm][b]{0.3cm}{$x$}} \llap{
		\parbox[][-1.15cm][b]{2.6cm}{$x_1$}} \llap{
		\parbox[][-1.15cm][b]{1.68cm}{$x_2$}}  \llap{
		\parbox[][-3cm][b]{3.1cm}{{\color{contourpink}$2\pi n$}}}
	\caption{Contour for calculating $\delta \log Z(n)$ in \eqref{toevaluate}.} \label{figContour} 
\end{figure}
\begin{figure}
	\centering \includegraphics[scale=0.16]{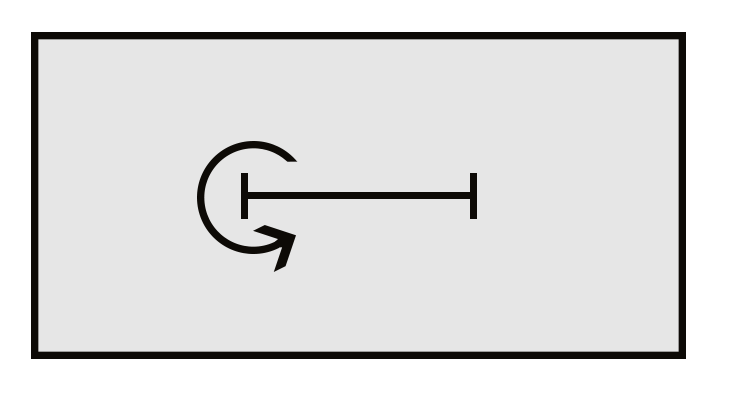} \llap{
		\parbox[][-2.1cm][b]{2.6cm}{$2\pi n$}} \llap{
		\parbox[][-0.5cm][b]{2.4cm}{$x_1$}} \llap{
		\parbox[][-0.5cm][b]{1.4cm}{$x_2$}}  \qquad \quad  \includegraphics[scale=0.16]{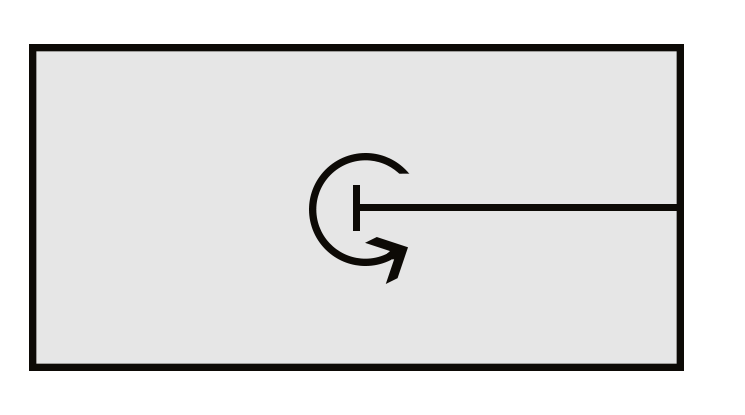} \llap{
		\parbox[][-2.1cm][b]{2.3cm}{$2\pi n$}} \llap{
		\parbox[][-0.5cm][b]{1.85cm}{$0$}} \llap{
		\parbox[][-0.5cm][b]{0.76cm}{$\infty$}} \qquad \quad  \includegraphics[scale=0.16]{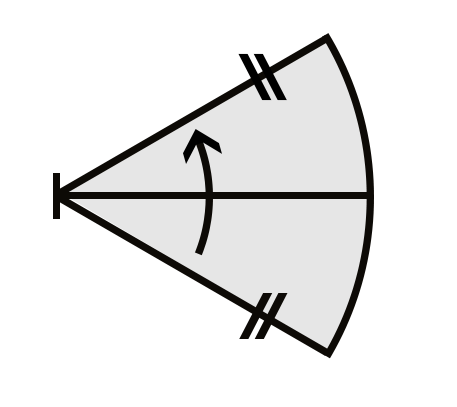} \llap{
		\parbox[][-1.8cm][b]{1cm}{$2\pi$}} \llap{
		\parbox[][-0.95cm][b]{1.9cm}{$0$}} \llap{
		\parbox[][-0.95cm][b]{0.76cm}{$\infty$}}
	\caption{Coordinate transformation \eqref{orbifoldcoord} from $\zeta$ on the replica manifold (left) to $\frac{\zeta-x_1}{x_2-\zeta}$ (middle) to $z$ on the orbifold, i.e.~the regular complex plane (right). The `pizza slice' representing the orbifold is drawn here for $n = 6$. 
	} \label{figPizza} 
\end{figure}

We can now consider a coordinate transformation that induces a change $\delta L$ of the length $L = x_2-x_1$ of the interval $A$, and then integrate $\delta \log Z(n)$ to find the dependence of $Z(n)$ on $x_1$ and $x_2$. We can choose for example the non-conformal transformation $x' = x + \theta(x-x_0)  \delta L$, where $x_1 < x_0 < x_2$. This indeed has the effect $L' = x_2'-x_1' = L + \delta L$. 
Then 
\ali{
	\delta \log Z(n) &= - \frac{\delta L}{2\pi} \int \vev{T_{xx}} \delta(x-x_0) dx dt_E = - \frac{\delta L}{2\pi} \int \vev{T_{xx}(x_0,t_E)} dt_E \\ 
	&= - \delta L \left( \oint_{x_1} \frac{d\zeta}{2\pi i} 
	\vev{T_{\zeta\zeta}(\zeta)} - \oint_{x_1} \frac{d\bar\zeta}{2\pi i} \vev{T_{\bar\zeta\bar\zeta}(\bar \zeta)} \right) \label{toevaluate}			  
} 
In the second line we moved to complex coordinates $\zeta = x + i t_E$, deforming the contour along the full infinite range of $t_E$ at $x=x_0$ to $\zeta$ encircling $x_1$ ($x_2$ is an equally good choice), see Fig \ref{figContour}. Now we still need to know $\vev{T_{\zeta\zeta}}$. On a regular complex plane with coordinate $z$, the stress tensor expectation value is zero because of rotational and translation invariance. The replica manifold can be conformally transformed into the complex plane, i.e.~uniformized, by the conformal transformation 
\ali{
	z = \left(\frac{\zeta-x_1}{x_2-\zeta}\right)^{1/n}  \label{orbifoldcoord} 
}
which consists of first mapping the branch points to $0$ and infinity, and then going to the $\mathbb{Z}_n$ orbifold, pictured as `pizza slice' in Fig \ref{figPizza}.     
Under this conformal transformation, the stress tensor of the CFT transforms anomalously, $T_{\zeta\zeta}(\zeta) = \left(\p z/\p \zeta\right)^2 T_{zz}(z) + \frac{c}{12} \{z,\zeta\}$, with $c$ the central charge of the CFT. It follows that $\vev{T_{\zeta\zeta}}$ is determined by the Schwarzian derivative of $z(\zeta)$ to be 
\ali{
	\vev{T_{\zeta\zeta}(\zeta)} = \frac{c}{24} \left(1 - \frac{1}{n^2}\right) \frac{(x_2-x_1)^2}{(\zeta-x_1)^2(\zeta-x_2)^2}. \label{Tzetazetavev} 
}  
In evaluating the complex integrals in \eqref{toevaluate}, the residue theorem picks up an extra $n$. This is because $\zeta$ is a complex coordinate with an argument of range $2\pi n$ around $x_1$ or $x_2$. 
It follows that 
\ali{
	\frac{\delta \log Z(n)}{\delta L} = - \frac{c}{6} \frac{n - 1/n}{L}   
}
or $\log Z(n) =  - \frac{c}{6} (n-1/n) \log L + ...$, where the dots are $L$-independent terms. 
We have obtained an expression for the $L$-dependence of $Z(n) = \tr \rho_A^\alpha|_{\alpha = n}$ for integer values of $\alpha=n\geq1$. 
The full partition function   
for complex $\alpha$ is then of the form $Z(\alpha) + \sin(\pi \alpha) g(\alpha)$ in the $\text{Re } \alpha > 1$ region with $g$ an analytic function. 
It can be shown based on the fact that $\hat \rho_A$ 
has eigenvalues $\lambda \in [0,1]$ and using Carlson's theorem \cite{Solodukhin:2011gn} 
that $g(\alpha) \equiv 0$ and therefore the obtained $Z(n)$ is valid beyond integer $n$. After substitution in \eqref{Sformula}, this leads to the famous formula for the vacuum entanglement of an interval $A$ in a 2-dimensional CFT  
\ali{
	S_A = \frac{c}{3} \log \frac{L}{\epsilon}. \label{Sresult}
}
Its scaling with $c$ shows the Weyl anomaly of the CFT crucially enters the derivation -- in \eqref{Tzetazetavev} in this case, as the Schwarzian derivative transformation rule is implied by the Weyl anomaly. 
The UV cutoff $\epsilon$ has to be introduced for dimensional reasons, and regulates the arbitrarily large contributions to the entropy from UV degrees of freedom arbitrarily close to $\p A$.

For the choice of $\delta L$-inducing coordinate transformation we could have also followed \cite{Holzhey:1994we} and chosen a scale transformation.

\subsection{Replicated target space} Let us also comment on the twist field correlator derivation of $\log Z(n)$, i.e.~the perspective of Fig \ref{figreplica}b. Details can be found in  \cite{Calabrese:2009qy,Calabrese:2004eu}. 

The stress tensor expectation value $\vev{T_{\zeta\zeta}(\zeta)}$ in \eqref{Tzetazetavev} corresponds to the insertion of one stress tensor operator on the replica manifold in Fig \ref{figreplica}a, let's say on the $i$-th sheet. From the point of view of Fig \ref{figreplica}b, it is the insertion of the stress tensor of the $i$-th copy of the theory in the presence of twist field operators and thus (normalizing left and right with the insertion of the unit operator) 
\ali{
	\left \langle T_{\zeta\zeta}(\zeta) \right \rangle_{{\mathcal L, \mathcal R_{n,A}}} \, = \, \frac{\left \langle T_{\zeta\zeta}^{(i)}(\zeta) \, T_n(x_1) \, \tilde T_n(x_2) \right \rangle_{\mathcal L^{(n)},\mathbb C}}{\left \langle T_n(x_1) \, \tilde T_n(x_2) \right \rangle_{\mathcal L^{(n)},\mathbb C}} , \label{condition}
}
where $x_1, x_2$ are short for the twist field locations $(\zeta_1,\bar \zeta_1), (\zeta_2,\bar \zeta_2)$ more generally, and  where we have included for clarity subscripts specifying the theory in which the expectation values are taken. 
The denominator of the right hand side is the $Z(n)$ we are to determine. For primary twist fields \cite{Cardy:2007mb} it takes the form $1/|x_1-x_2|^{2 d_n}$, with $d_n$ the unknown scaling dimension of $T_n$ and $\tilde T_n$.   
The numerator of the right hand side can then be rewritten in terms of the twist field two-point function,  
\ali{
	\left \langle T_{\zeta\zeta}^{(i)}(\zeta) \, T_n(x_1) \, \tilde T_n(x_2) \right \rangle = \frac{1}{n} \frac{d_n}{2} \frac{(x_1-x_2)^2}{(\zeta-x_1)^2 (x_2 - \zeta)^2} \left \langle T_n(x_1) \, \tilde T_n(x_2) \right \rangle, \label{fromWardid}
}
by applying the Ward identity for the $(\mathcal L^{(n)}, \mathbb{C})$ theory with stress tensor $T_{\zeta\zeta}^{(n)} = n T_{\zeta\zeta}^{(i)}$ (by extensivity of the Lagrangian and thus Hamiltonian density). 
Then \eqref{Tzetazetavev}, \eqref{condition} and \eqref{fromWardid} impose $d_n = \frac{c}{12}(n - 1/n)$, leading to the same $\log Z(n)$ as in the previous section, and the same result for $S_A$ in \eqref{Sresult}. 

One main advantage of this perspective is that the known transformation behavior of the 
conformal two-point function \eqref{TSZnbis} under conformal transformations immediately gives us the formula for $\log Z(n)$ and thus the entanglement $S_A$ in a conformally related geometry, as we will now briefly discuss.  

In the notation $f = x + i t_E, \bar f=x-i t_E$ for our current set-up, with metric $ds^2 = df d\bar f = dt_E^2 + dx^2$ and state the vacuum state $|0\rangle_f$ as measured in $f$-coordinates, 
we have   
\ali{
	S_A = \frac{c}{3} \log \left| \frac{f_1 - f_2}{\epsilon_f} \right| \label{planeformula}
}
for 
\ali{
	\langle \, T_n(f_1,\bar f_1) \tilde T_n(f_2, \bar f_2) \, \rangle = |f_1-f_2|^{-\frac{c}{6}(n - 1/n)}.  
}
Under a conformal transformation 
$f = f(z)$, $\bar f = \bar f(\bar z)$, the metric transforms $ds^2 = \left|\p f/\p z \right|^2 dz d\bar z \equiv \Omega(z,\bar z) dz d\bar z$ with a Weyl factor $\Omega$, and the entanglement  
in $z$-coordinates becomes 
\ali{
	S_A = \frac{c}{6} \log  \frac{(f(z_1) - f(z_2))(\bar f(\bar z_1) - \bar f(\bar z_2))}{\sqrt{f'(z_1) f'(z_2) \bar f'(\bar z_1) \bar f'(\bar z_2)} \epsilon_z \epsilon_{\bar z}}. \label{Sconf}
}
This follows from 
\ali{
	\langle \, T_n(z_1,\bar z_1) \tilde T_n(z_2, \bar z_2) \,\rangle &= \left|\frac{f(z_1)-f(z_2)}{\sqrt{f'(z_1)f'(z_2)}}\right|^{-\frac{c}{6}(n - 1/n)} \\ 
	&= |f'(z_1) f'(z_2)|^{d_n}  \langle \, T_n(f_1,\bar f_1) \tilde T_n(f_2, \bar f_2) \, \rangle  
} 
where we used the transformation behavior of a primary field \\
$\mathcal O(z,\bar z) =\left(\p z'/\p z\right)^h \left(\p \bar z'/\p \bar z\right)^{\bar h} \mathcal O'(z',\bar z')$ 
for the twist fields of dimension $h = \bar h = d_n/2$ under an 
$f(z)$ transformation.

As can be seen from comparison of \eqref{planeformula} and \eqref{Sconf}, the UV cutoffs as measured in $f$ or $z$ coordinates are related by \cite{Holzhey:1994we} 
\ali{
	\epsilon_f = f'(z) \epsilon_z   \label{cutoffs}
}
(and writing $\epsilon$ as $\sqrt{\epsilon_1 \epsilon_2}$). 
Indeed, from $f(z+\epsilon_z) \approx f(z) + \epsilon_z f'(z)$ with $\epsilon_z \equiv \delta z$, it follows that $\epsilon_f \equiv \delta f$ is given by the above.  

\begin{figure}
	\qquad \qquad \qquad 
	\includegraphics[scale=0.115]{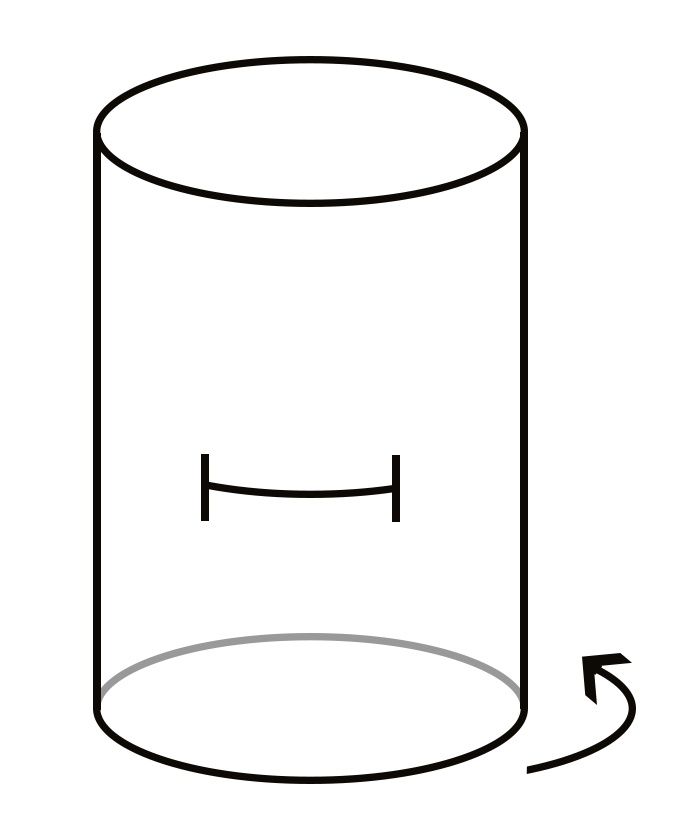} \llap{
		\parbox[][-2.1cm][b]{1.75cm}{$u$}} \llap{
		\parbox[][-2.1cm][b]{1.1cm}{$v$}} \llap{
		\parbox[][-0.3cm][b]{0.25cm}{$\phi$}}
	\qquad \qquad \qquad 
	\includegraphics[scale=0.115]{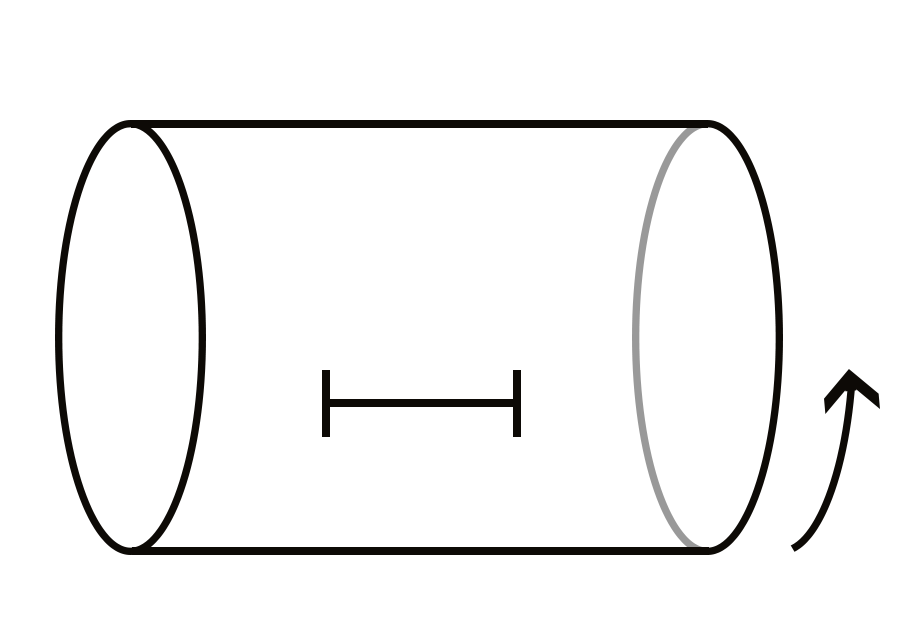} \llap{
		\parbox[][-1.7cm][b]{1.95cm}{$u$}} \llap{
		\parbox[][-1.7cm][b]{1.3cm}{$v$}} \llap{
		\parbox[][-0.3cm][b]{0.25cm}{$\tau$}}
	\caption{
		Left:  Finite size formula $S_A = \frac{c}{3} \log \left( \frac{\Sigma}{\pi \epsilon} \sin \frac{\pi (v-u)}{\Sigma} \right)$ for an interval $z_1 = \bar z_1 = u$, $z_2 = \bar z_2 = v$ on the cylinder of size $\Delta \phi = \Sigma$ follows 
		from \eqref{Sconf} with $f(z) = \exp{ \left( \frac{2\pi}{\Sigma} i z \right)}$ relating the $f$-coordinate of the plane to the $z$-coordinate of the cylinder with compact $\phi = \text{Re }z$.  
		Right:  Finite temperature formula $S_A = \frac{c}{3} \log \left( \frac{\beta}{\pi \epsilon} \sinh \frac{\pi (v-u)}{\beta} \right)$ for an interval $z_1 = \bar z_1 = u$, $z_2 = \bar z_2 = v$ on the cylinder of size $\Delta \tau = \beta$ follows from \eqref{Sconf} with $f(z) = \exp{ \left(\frac{2\pi}{\beta} z\right)}$ relating the $f$-coordinate of the plane to the $z$-coordinate of the cylinder with compact $\tau = \text{Im }z$.   
	}
	\label{figcylS}
\end{figure}

The formula \eqref{Sconf} can be applied immediately to the cases pictured in Fig \ref{figcylS} of intervals in the $z$-cylinder.  

\section{Relation to thermal entropy} \label{sectionThermal}

There is a `short-cut' for deriving the interval entanglement $S_A$ in \eqref{Sresult} which will also be important for the holographic interpretation. It consists of mapping the set-up of section \ref{sectionreplicaCFT} to a thermal system. This section follows \cite{Holzhey:1994we}.   

A conformal change of coordinates induces a change of basis among the operators of the theory and affects the density matrix 
through a unitary transformation, to which the trace in \eqref{SAdef} 
is insensitive. 
Moreover, the anomalous ($\sim c$) contribution to the transformed Hamiltonian only affects the unnormalized density matrix (and thus e.g.~$\log(Z)$), but not the normalized density matrix that appears in the von Neumann entropy \eqref{SAdef}. 
This is to say that the entanglement entropy is conformal invariant, and hence its calculation can be simplified by considering well-chosen conformal mappings.

\begin{figure}
	\centering \includegraphics[scale=0.18]{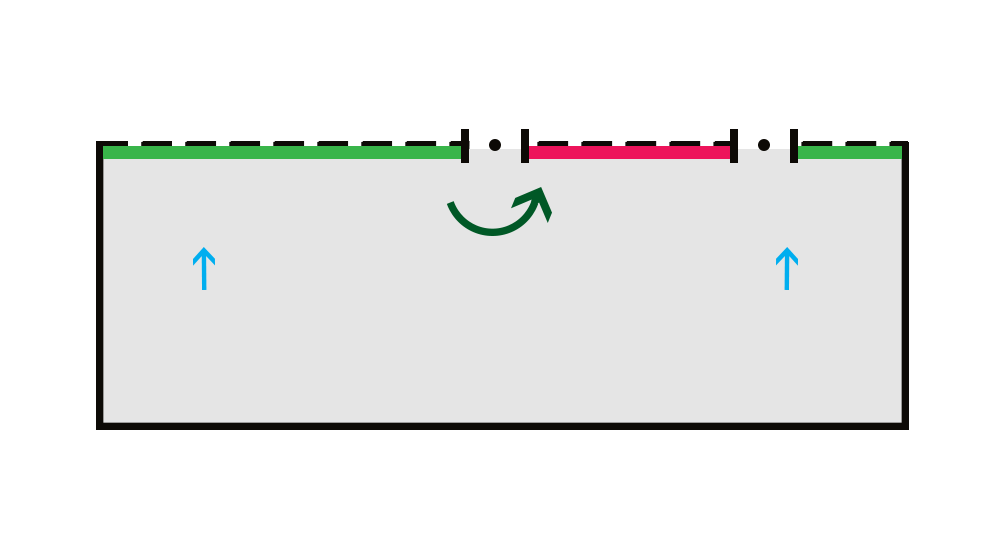}\llap{
		\parbox[b]{5cm}{a)\\\rule{0ex}{2cm}}} \llap{
		\parbox[][-4.1cm][b]{2.45cm}{$\epsilon$}} \llap{
		\parbox[][-4.1cm][b]{1.8cm}{$L-\epsilon$}} \llap{
		\parbox[][-2.3cm][b]{2.7cm}{{\color{taugreen} $\tau$}}} 
	\qquad \qquad \qquad \includegraphics[scale=0.18]{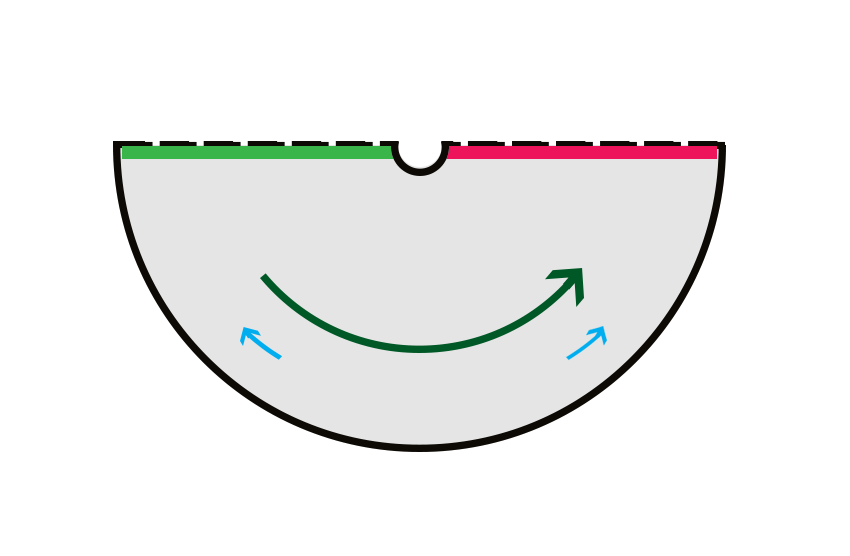}\llap{
		\parbox[b]{5cm}{b)\\\rule{0ex}{2cm}}} \llap{
		\parbox[][-4.1cm][b]{2.1cm}{$\frac{\epsilon}{L}$}} \llap{
		\parbox[][-4.1cm][b]{0.9cm}{$\frac{L}{\epsilon}$}} \llap{
		\parbox[][-2.3cm][b]{2.4cm}{{\color{taugreen} $\tau$}}} \\ 
	\quad \includegraphics[scale=0.18]{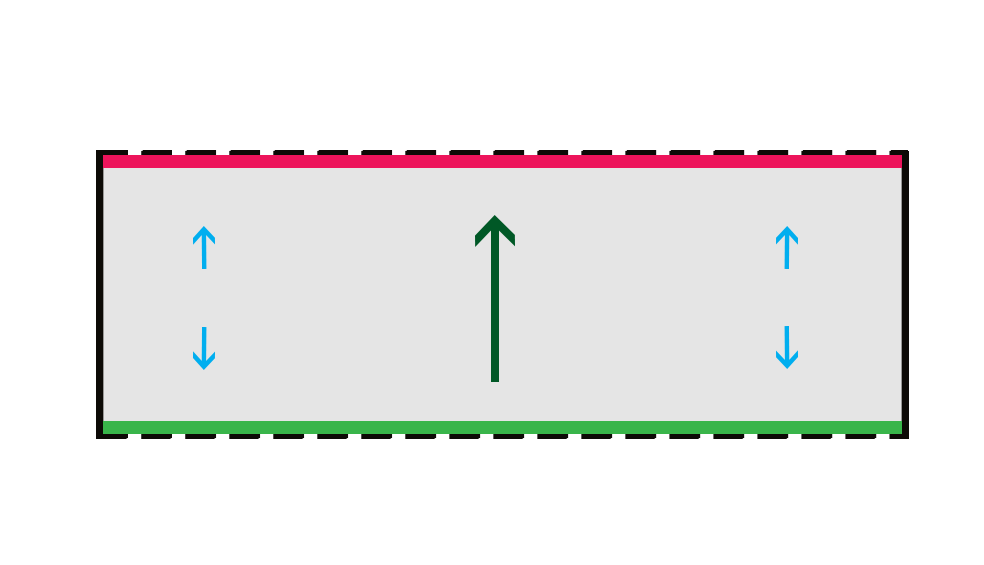}\llap{
		\parbox[b]{5.4cm}{c)\\\rule{0ex}{2.5cm}}}  \llap{
		\parbox[][-4.1cm][b]{4.6cm}{$\frac{1}{\kappa}\log\frac{\epsilon}{L}$}} \llap{
		\parbox[][-4.1cm][b]{0.9cm}{$\frac{1}{\kappa}\log\frac{L}{\epsilon}$}} \llap{
		\parbox[][-3.4cm][b]{5.1cm}{$0$}} \llap{
		\parbox[][-.9cm][b]{5.5cm}{$-\frac{i \pi}{\kappa}$}} \llap{
		\parbox[][-2.3cm][b]{2.5cm}{{\color{taugreen} $\text{Im } z = \tau$}}} 
	\qquad \qquad 
	\includegraphics[scale=0.18]{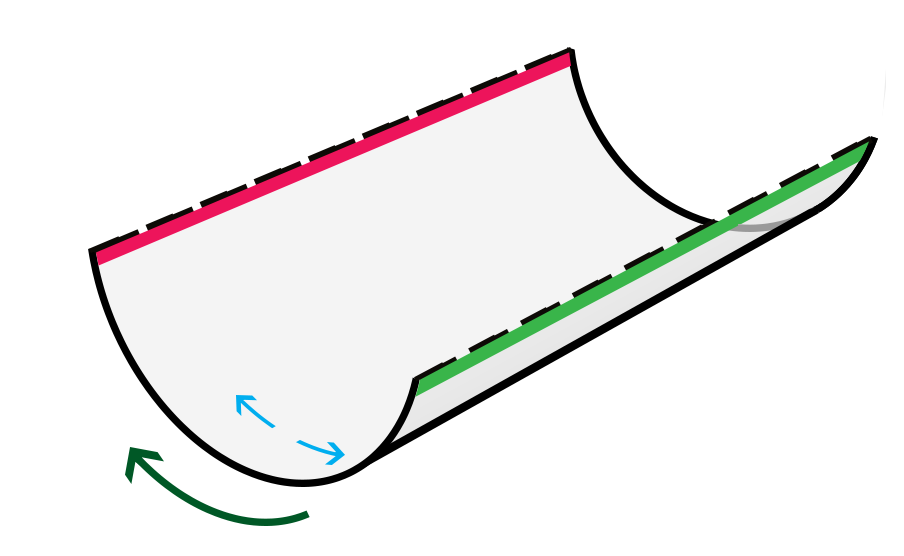} \llap{
		\parbox[][-2.4cm][b]{5.1cm}{or}} \llap{
		\parbox[][-1.2cm][b]{4.2cm}{{\color{taugreen} $\tau$}}}
	\caption{ 
		Prepared state, with region $A$ in red and complementary region $\bar A$ in green, under conformal mappings between the $\zeta$, $w$- and $z$-geometry, in respectively a), b) and c). 
		In terms of metrics (for $\kappa=1$ choice), $d\tau^2 + \frac{dR^2}{R^2} = dz d\bar z$ (cylinder)  
		$=d(\log w) d(\log \bar w) = \frac{1}{w \bar w} dw d\bar w$ 
		$ \ra dw d\bar w$ (plane) $= R^2 d\tau^2 + dR^2$ with $w = R \exp(i \tau)$ and $z = x_z + i \tau = (\log R) + i \tau$. Under these mappings, a periodic coordinate $\tau$ around $\p A$ becomes the angle on the annulus or the Euclidean time coordinate of the cylinder. The small blue arrows point in the direction of the Euclidean evolution $t_E$.   
	} \label{figHolzheypsi}
\end{figure}

\begin{figure}
	\centering \includegraphics[scale=0.16]{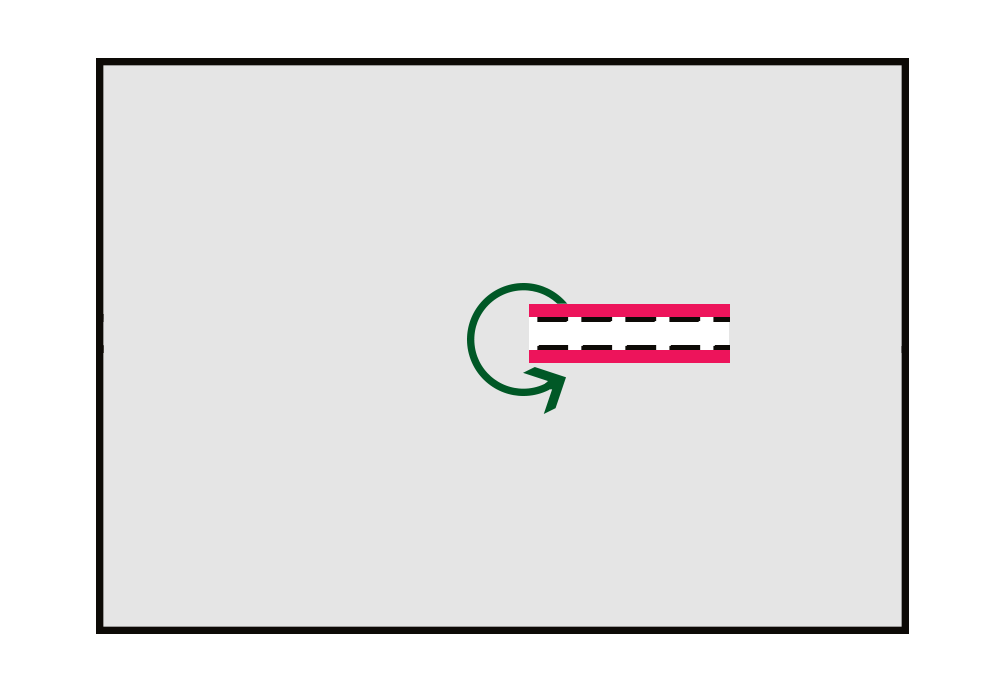}\llap{
		\parbox[b]{5cm}{a)\\\rule{0ex}{3cm}}} \llap{
		\parbox[][-2.7cm][b]{2.6cm}{${\color{taugreen} \tau}$}}  \llap{
		\parbox[][-2cm][b]{1.9cm}{$(\phi)$}} \llap{
		\parbox[][-3.4cm][b]{2cm}{$(\phi')$}} \qquad \qquad \qquad \includegraphics[scale=0.16]{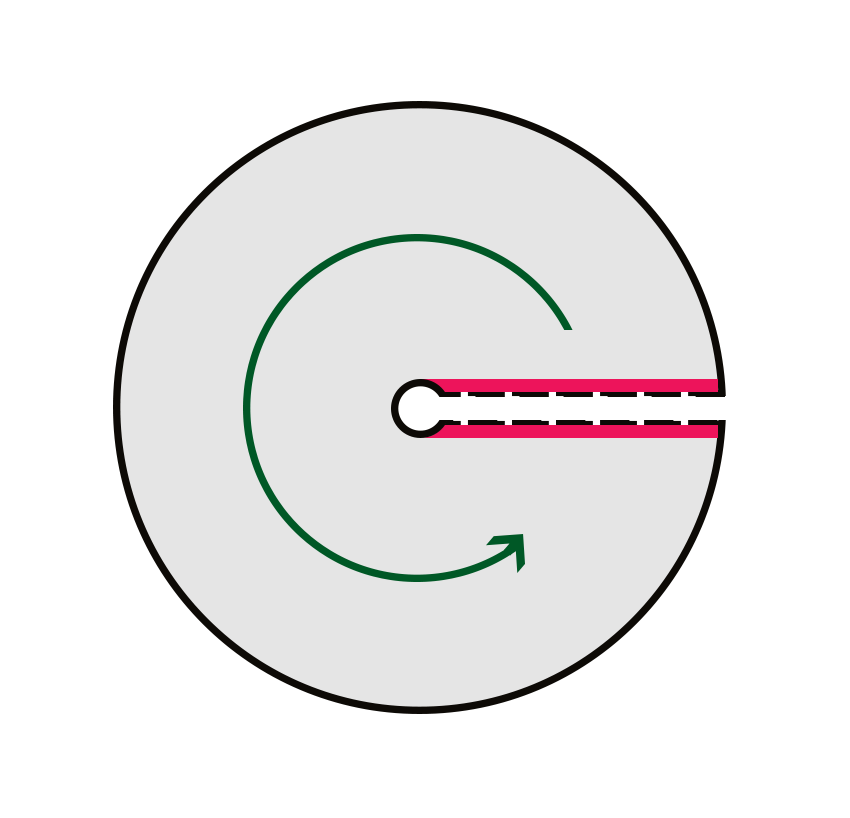}\llap{
		\parbox[b]{5cm}{b)\\\rule{0ex}{3cm}}} \llap{
		\parbox[][-3cm][b]{2.9cm}{${\color{taugreen} \tau}$}} \llap{
		\parbox[][-2.5cm][b]{1.3cm}{$(\phi)$}} \llap{
		\parbox[][-3.7cm][b]{1.4cm}{$(\phi')$}} \\ 
	\quad \includegraphics[scale=0.16]{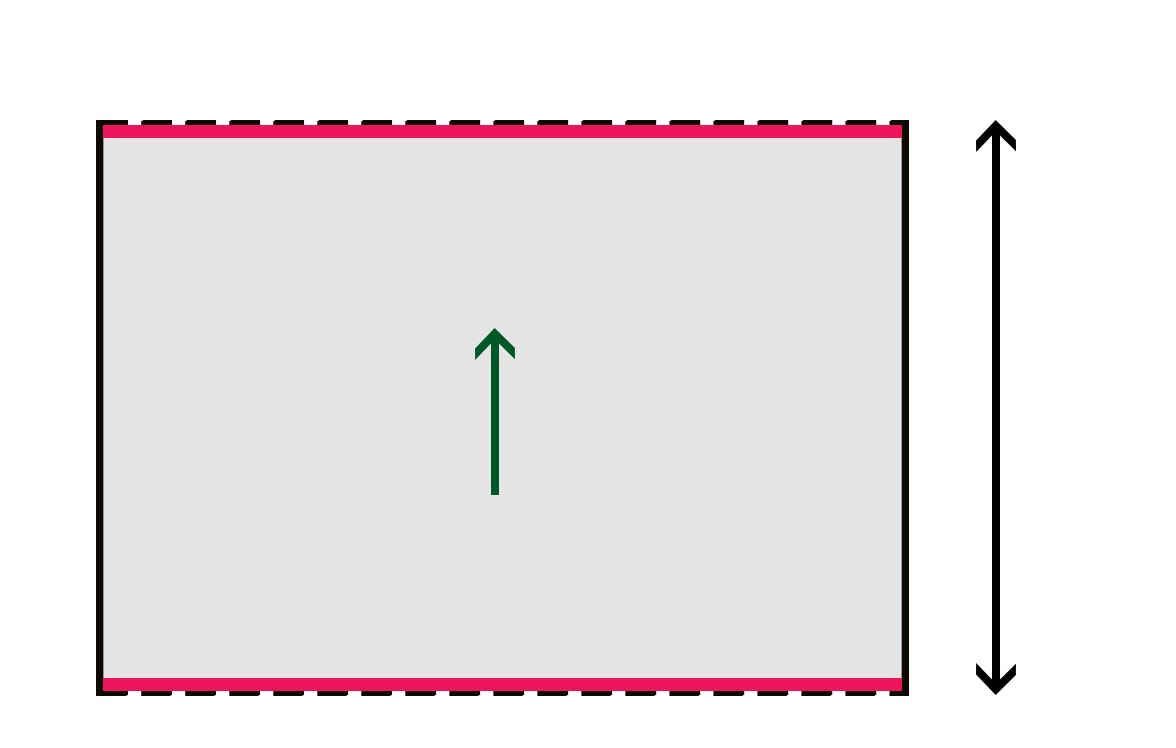}\llap{
		\parbox[b]{5.4cm}{c)\\\rule{0ex}{2.5cm}}} \llap{
		\parbox[][-2.3cm][b]{2.7cm}{${\color{taugreen} \tau}$}} \llap{
		\parbox[][-5.35cm][b]{3.3cm}{$(\phi)$}} \llap{
		\parbox[][0.35cm][b]{3.4cm}{$(\phi')$}} \llap{
		\parbox[][-2.3cm][b]{0.6cm}{$\frac{2\pi}{\kappa}$ or $\beta$}} 
	\quad \quad  
	\includegraphics[scale=0.17]{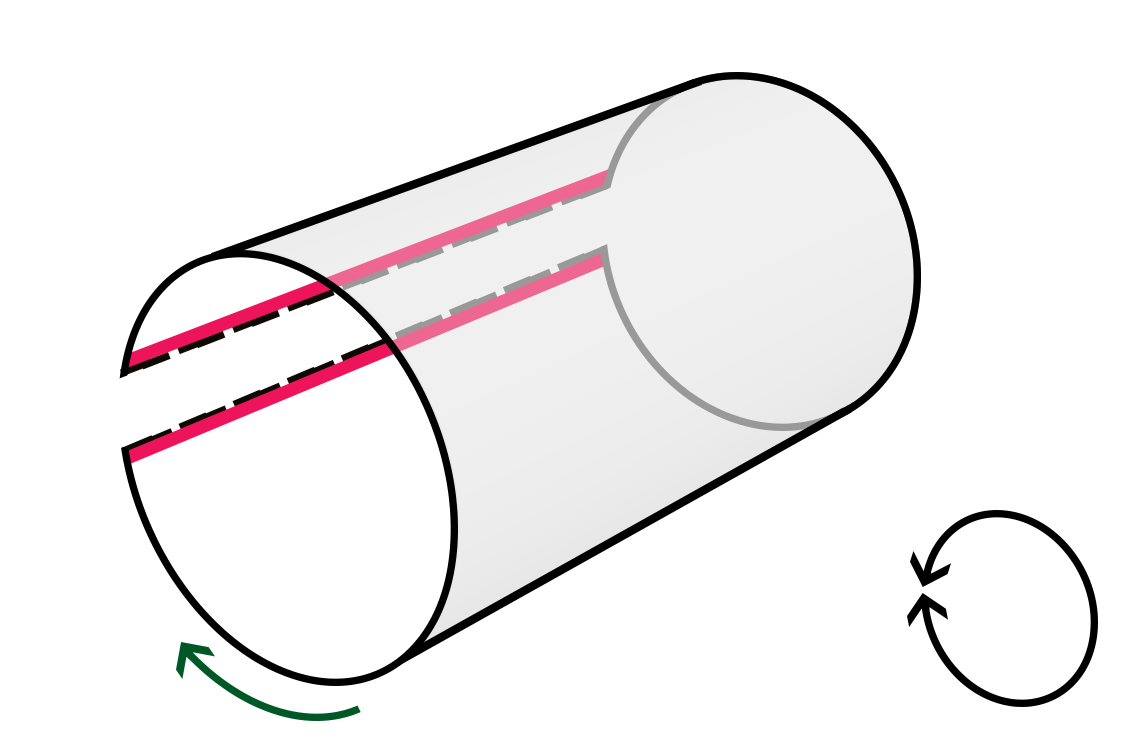} \llap{
		\parbox[][-2.4cm][b]{5.5cm}{or}} \llap{
		\parbox[][-1.2cm][b]{4.9cm}{{\color{taugreen} $\tau$}}} \llap{
		\parbox[][-4.65cm][b]{3.3cm}{$(\phi')$}} \llap{
		\parbox[][-3.1cm][b]{3.4cm}{$(\phi)$}} \llap{
		\parbox[][-1.2cm][b]{1.65cm}{$\beta$}} 
	\caption{
		Reduced density matrix $\rho_A$ for the state in Fig \ref{figHolzheypsi}, under conformal mappings between the $\zeta$, $w$- and $z$-geometry, in respectively a), b) and c). In brackets are the field configurations to read off the matrix elements $(\rho_A)_{\phi \phi'} = \langle \phi| \rho_A | \phi'\rangle$.    		
	} \label{figHolzheyrhoA} 
\end{figure}

We consider the same theory $Z \equiv Z(1)$ of section \ref{sectionreplicaCFT}, in complex coordinates $\zeta = x + i t_E$, $\bar \zeta = x - i t_E$. First we translate the interval $A$ from $x = [x_1, x_2]$ to $x = [0,L]$. The vacuum state of the system is pictured in Fig \ref{figHolzheypsi}a.   Then we consider the conformal transformation to $w = (\zeta - x_1)/(x_2-\zeta)$ or 
\ali{
	w = \frac{\zeta}{L-\zeta},  
}
which maps the interval to the positive half-line. 
Keeping track of the UV cutoff, the more precise statement is that the $x = [\epsilon, L-\epsilon]$ interval is mapped to $w = \text{Re }w = [\frac{\epsilon}{L} ,\frac{L}{\epsilon}]$. Then, we further transform to 
\ali{ 
	z = \frac{1}{\kappa} \log w, 
} 
with $\kappa$ an arbitrary real number, that should therefore not affect  
any physics.  
It maps the positive half-line to $z = \text{Re }z = [\frac{1}{\kappa} \log \frac{\epsilon}{L} , \frac{1}{\kappa} \log \frac{L}{\epsilon}]$ and the negative half-line to $z = [\frac{1}{\kappa} \log \frac{\epsilon}{L} - \frac{i \pi}{\kappa}, \frac{1}{\kappa} \log \frac{L}{\epsilon} - \frac{i \pi}{\kappa}]$, so that $A$ is now one side of a strip of width $\pi/\kappa$. The partition function $Z(1) = \tr \rho$ 
is mapped to the $w$-\emph{annulus} and the $z$-\emph{cylinder} respectively. The imaginary part of $z$ 
\ali{
	\text{Im }z \equiv \tau 
}
brings forth a periodic coordinate; as the angle of the $w$-annulus with periodicity $\Delta \tau = 2\pi$, and as the compact direction of the $z$-cylinder with periodicity $\Delta \tau = 2\pi/\kappa$. 
A direct consequence of this is that the reduced density matrix $\rho_A$ in both pictures takes the form of, not just any mixed density matrix, but a \emph{thermal} density matrix 
\ali{ \rho_A= e^{-(\Delta \tau) H_\tau} = \rho_{thermal} \label{BWtheorem}} 
as demonstrated in Fig \ref{figHolzheyrhoA}.  
On the annulus, $H_\tau$ is the generator of rotation in the Euclidean plane, and thus the boost generator in Lorentzian signature. On the cylinder, $H_\tau$ is the actual Hamiltonian or generator of cylinder (Euclidean) time translation. 
The corresponding inverse temperature is denoted $\beta$ by choosing $\kappa = 2\pi/\beta$. It is a fictitious temperature measured by an observer that only has access to half of the $w$-plane or only one side of the $z$-strip. 
Fig \ref{figHolzheyrhoA}b gives us the path integral derivation of the Bisognano Wichmann theorem  \eqref{BWtheorem} \cite{Bisognano:1976za}  (e.g.~\cite{UnruhWeiss:1983ac,Headrick:2019eth}). 

In the cylinder picture, $S_A$ is now equal to the thermal entropy $S_{thermal}$ of the strip of width $\beta$, which is simply given by the thermodynamic formula $S_{thermal} = \beta E + \log Z(\beta)$  
in terms of the energy and free energy, or  
\ali{
	S_{thermal} = (1 - \beta \p_\beta) \log Z(\beta).  \label{eqSthermal}
}
The partition function of the CFT on the strip of width $\beta$ and length $L_z \gg \beta$ was determined in \cite{Bloete:1986qm,Affleck:1986bv},  
again by making use of the Weyl anomaly, to be (to leading order in $1/\beta$) 
\ali{
	\log Z(\beta) = \text{const } \beta \,  L_z + \frac{\pi c}{6 \beta} L_z \, .   
}
The constant in the first term is non-universal, i.e.~CFT-dependent, and drops out of the formula for $S_{thermal}$, which gives 
\ali{
	S_{thermal} = \frac{c \pi}{3 \beta} L_z = \frac{c}{3} \log\frac{L}{\epsilon} \, . 
}
$L_z$ is the full (divergent) length $A$ of the cylinder in Fig \ref{figHolzheypsi}c, which is given by $\frac{\beta}{\pi} \log \frac{L}{\epsilon} \equiv \frac{\beta}{\pi} \log \frac{L_\zeta}{\epsilon_\zeta}$, so that in the second equality we find, as we should, $S_A = \frac{c}{3} \log\frac{L}{\epsilon}$ in \eqref{Sresult}.  

As pointed out in \cite{Casini:2011kv}, there is an IR/UV quality to the relation $L_z \sim \log L/\epsilon$ between the cylinder regulator (large $L_z$) and the plane regulator (small $\epsilon$), which you typically find in a holographic setting\footnote{ 
	The difference is in the contractability of the $\tau$-circle (non-contractible on the cylinder, contractible on the plane), 
	which is also reflected in the holographic description (non-contractible on the cylinder, contractible in the bulk description, see Fig \ref{figBTZ}).  
	}.   
A last comment about the derivation is that the formula for the entanglement of a finite interval on the thermal cylinder in Fig \ref{figcylS}b should become the thermal entropy for the state in Fig \ref{figHolzheypsi}c in the limit of $L_z \gg \beta$. This can be easily checked: $\lim_{L_z \ra \infty} \log(\beta/(\pi \epsilon_z) \sinh (\pi L_z/\beta)) = \pi L_z /\beta + \log (\beta/(2\pi \epsilon_z))$, with the second term vanishing for $\epsilon_z = \beta/(2\pi) \ll L_z$ following from \eqref{cutoffs} with $z = \beta/(2\pi) \log (\zeta/(L-\zeta))$.

\begin{figure}
	\centering \includegraphics[scale=0.12]{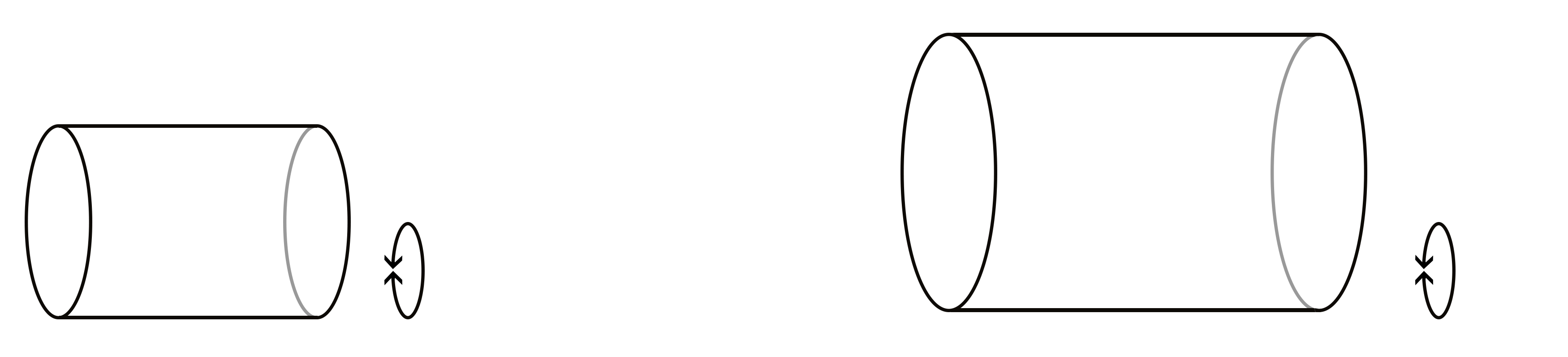} \llap{
		\parbox[][-0.4cm][b]{8cm}{$\beta$}} \llap{
		\parbox[][-0.4cm][b]{0.8cm}{${n \beta}$}}
	\caption{$Z(1)$ and $Z(n)$} 
	\label{figZncylCFT}
\end{figure}

In the above paragraph, $Z(\beta)$ is the $Z(1)$ on the thermal cylinder (where $\beta$ could also be taken to be $2\pi$ by choosing $\kappa=1$). 
The replicated manifold $\tr \rho_A^n$, obtained by gluing $n$ copies of 
the $Z(1)$ strip along $A$, is again a cylinder, $Z(n \beta) \equiv Z(n)$, only this time with periodicity $n \beta$ (Fig \ref{figZncylCFT}). By the conformal mapping to the cylinder, the conical singularities at $\p A$ have been removed from the replica manifold, and the analytic continuation to non-integer $n$ is immediate, since the periodicity of the cylinder can be varied continuously.    
The replica trick \eqref{Sformula} indeed 
gives $S_A = S_{thermal}$ \ali{
	S_{thermal} = (1 - n \p_n) \log Z(n)|_{n \ra 1} = \frac{c \pi}{3 \beta} L_z 
}
from the trivial $n$-dependence in $Z(n) \equiv Z(n \beta) = \frac{\pi c}{6 n \beta} L_z + \text{const } n \beta L_z$.  
The triviality of the thermal replica trick on the cylinder, combined with the argument that $S_A$ is conformal invariant, signals that the replica derivation of the interval entanglement $S_A$ and in particular the question of analytic continuation of $n$ should be well-defined (section \ref{sectionreplicaCFT}), as should its holographic interpretation (section \ref{sectholoholzhey}).  

The replica manifold in the annulus picture does have a conical 
singularity at the origin $\epsilon \ra 0$ with conical excess $2\pi (n-1)$. The length of the region $A$ can be varied by a scale transformation $x^\mu \ra x'^\mu = (1 - 2 \frac{\delta\epsilon}{\epsilon}) x^\mu$, such that $Z(n)$ in \eqref{deltalogZn} has to satisfy  
\ali{
	\delta \log Z(n) = \frac{\delta \log \epsilon}{\pi} \int \vev{T^\mu_{\phantom{\mu}\mu}} d^2 x . 
}
Here, the trace of the stress tensor can be obtained from applying the argument leading to \eqref{Tzetazetavev}, or by directly making use of the Weyl anomaly $\vev{T^\mu_{\phantom{\mu}\mu}} = \frac{c}{12} R$ 
which relates it to the curvature $R \sim (1-n) (\delta^{(2)}(x_1) + \delta^{(2)}(x_2))$ of the manifold and its boundaries (see also \cite{Ryu:2006ef} e.g.). This procedure for deriving \eqref{Sresult}, detailed in \cite{Holzhey:1994we}, gives an alternative to the replica trick derivation of subsection \ref{subsectreplica} that makes the coarse graining physics behind the geometric entropy $S_A$ apparent.

\section{Thermofield double}  \label{sectTFD}

In the previous section we encountered the  
possibility of an observer restricted to a part of spacetime $A$ measuring a state $\rho_A$ that is thermal even though the full system is in the vacuum state. This fits in the general concept of the thermofield double 
\cite{Takahasi:1974zn,Martin:1959jp}, which will be discussed in this section.

A thermofield double (TFD) is a particular vacuum state that is constructed to reproduce thermal physics of a given QFT. We will focus on a conformal QFT in particular, with a Hamiltonian $H_R$ and Hilbert space $\mathcal H_R$, and on $1+1$ dimensions. 
The Euclidean path integral for compactified Euclidean time with period $\Delta \tau = \beta$ gives the thermal partition function $Z(\beta)$ of the theory. In path integral visualization, 
$Z(\beta) = \tr \rho_{thermal}$ is a cylinder $\mathcal C(\beta)$ when the spacelike direction of the theory is non-compact, and thus $\rho_{thermal}$ is the cylinder before tracing, i.e.~with open cut  
\ali{
	\rho_{thermal} = \adjincludegraphics[width=2cm,valign=c]{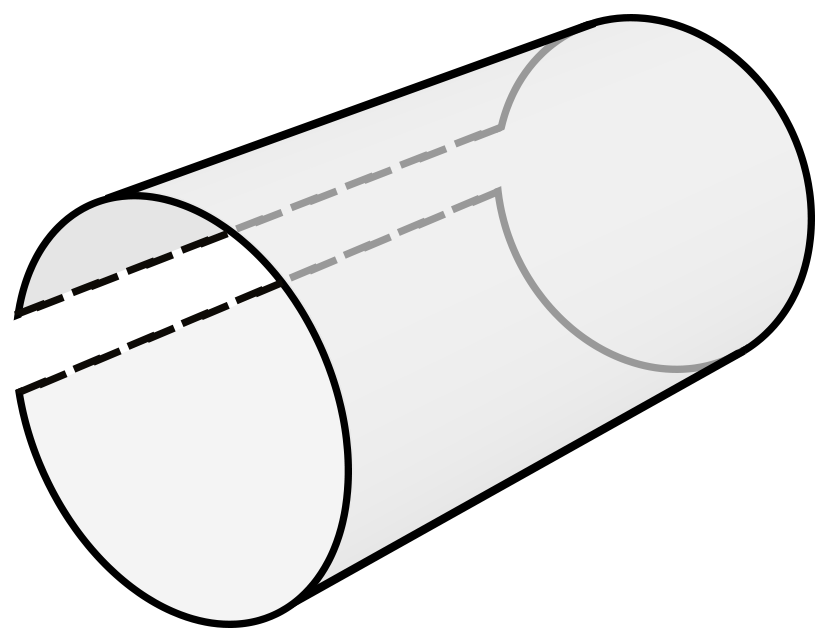}.  
} 
The statement is that the same physics can be described by constructing a state $|\psi\rangle$, called the TFD state, as the state 
\ali{
	|\psi\rangle = \adjincludegraphics[width=2.5cm,valign=c]{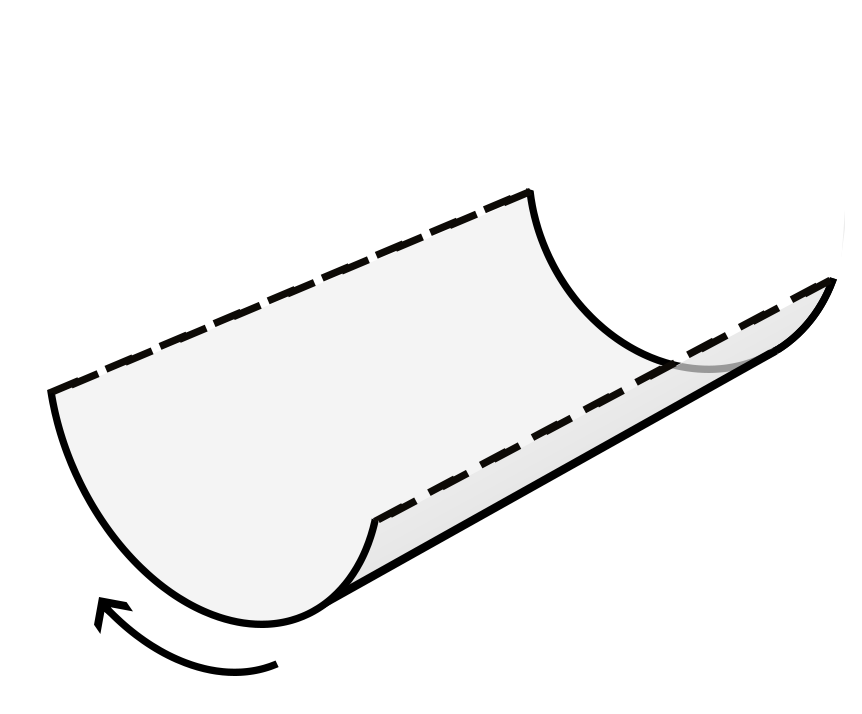} \llap{
		\parbox[][1.3cm][b]{2.5cm}{{\scriptsize$\tau$}}} \llap{
		\parbox[][2.3cm][b]{2.15cm}{{\tiny $\beta/2$}}} \llap{
		\parbox[][-0.7cm][b]{2.1cm}{{\scriptsize$(\phi_R)$}}} \llap{
		\parbox[][-0.4cm][b]{0.8cm}{{\scriptsize$(\phi_L)$}}}   
	\label{TFDpicture}
}
living in the doubled Hilbert space $\mathcal H_L \times \mathcal H_R$, with total Hamiltonian $H = H_R-H_L$ (as time $\tau$ runs downwards in the left copy and upwards in the right copy). We refer to the left copy as the one where $\tau$ starts and the right copy where $\tau$ arrives. This state is constructed such that it satisfies the property that its density matrix 
\ali{
	\rho = |\psi\rangle \langle \psi| =  \adjincludegraphics[width=2cm,valign=c]{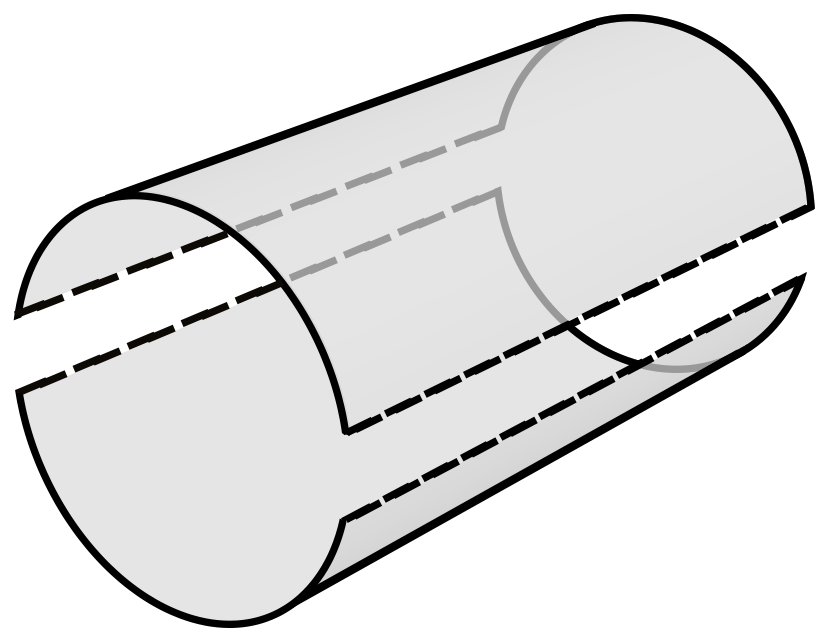} \label{rhoTFD}
}
reduces to the thermal density matrix of the original `right' system when the left copy is traced out\footnote{
	In equation \eqref{rhoATFD} and in the rest of the section, the pictorial representation of tracing out fields is short for the notation introduced in \eqref{trrhoblue}, leaving out the explicit integration over the fields for conciseness. 
}, 
\ali{
	\rho_R \equiv \tr_L \rho = \adjincludegraphics[width=2cm,valign=c]{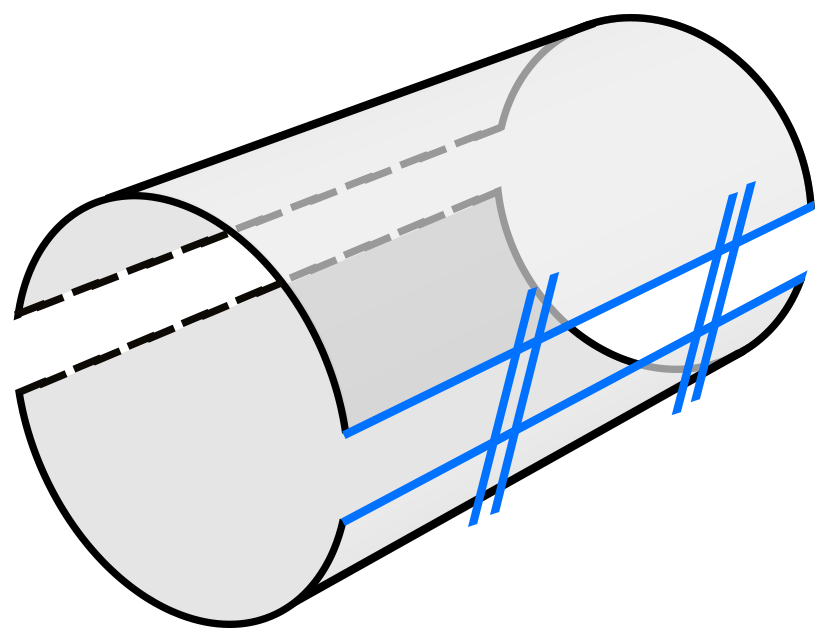} =  \adjincludegraphics[width=2cm,valign=c]{figs/TFDrhothermal.png} = \rho_{thermal}. 
\label{rhoATFD}   
}

Said otherwise, the TFD state is the answer to the question \cite{Laflamme:1988wg}  ``is there a state $|\psi\rangle \equiv |0(\beta)\rangle$  
for which the QFT vacuum expectation value $\langle \psi|\mathcal O|\psi\rangle = \tr(\mathcal O \rho)$ 
reproduces the statistical average $\tr(\mathcal O \rho_{thermal})$ 
of an operator $\mathcal O$?", 
\ali{
	\langle	\psi|\mathcal O|\psi\rangle &= \tr(\mathcal O \rho_{thermal}) \label{TFDstatement1} \\
	\adjincludegraphics[width=2cm,valign=c]{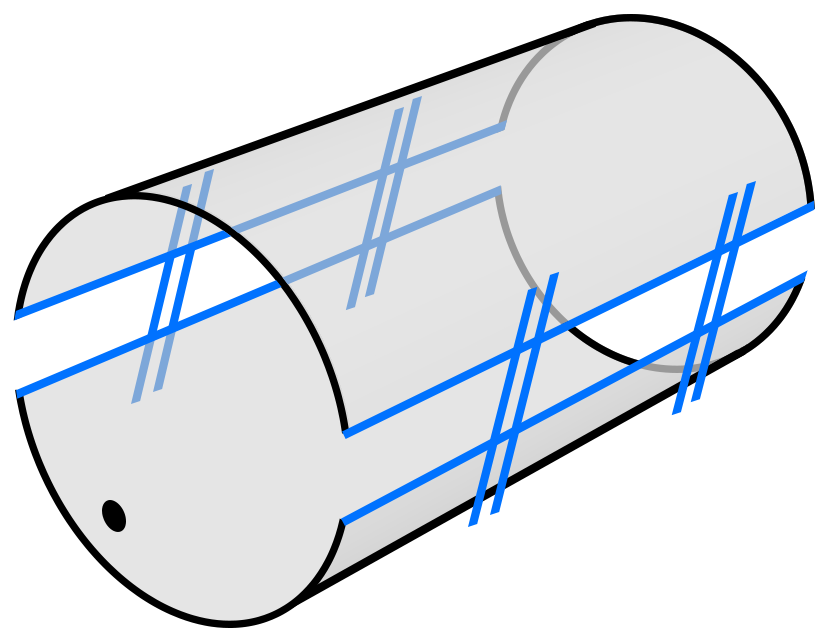}\llap{
		\parbox[][1.6cm][b]{2.1cm}{$\mathcal O$}} &=  \adjincludegraphics[width=2cm,valign=c]{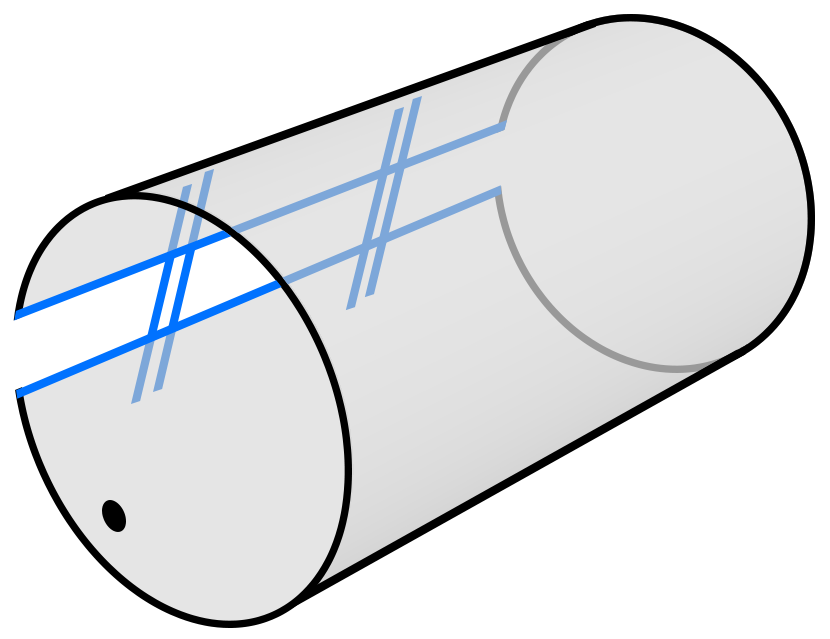}\llap{
		\parbox[][1.6cm][b]{2.1cm}{$\mathcal O$}}. 
}
Of course, for $\mathcal O$ the identity operator, this is just the statement that 
\ali{
	\langle \psi|\psi \rangle &= \tr(\rho_{thermal}) \label{} \\ 
	\adjincludegraphics[width=2cm,valign=c]{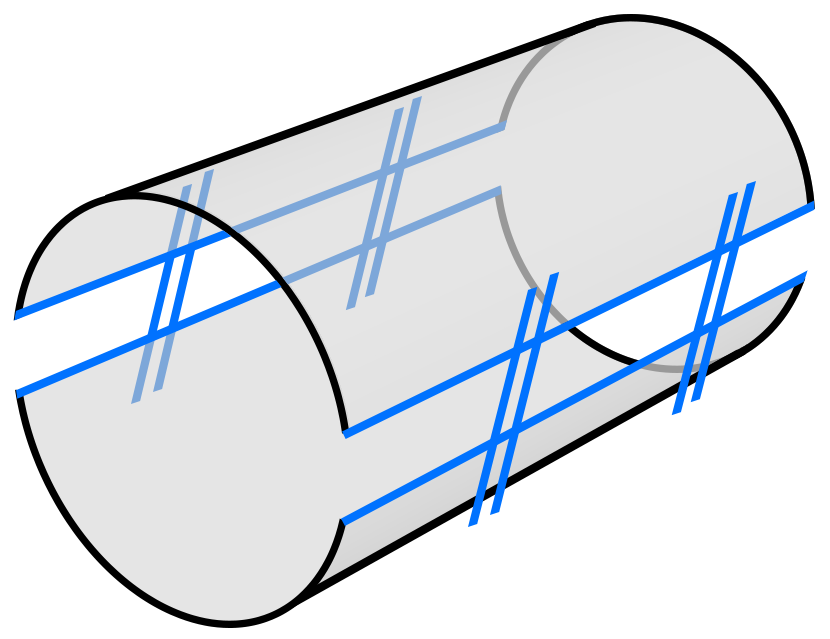} &=  \adjincludegraphics[width=2cm,valign=c]{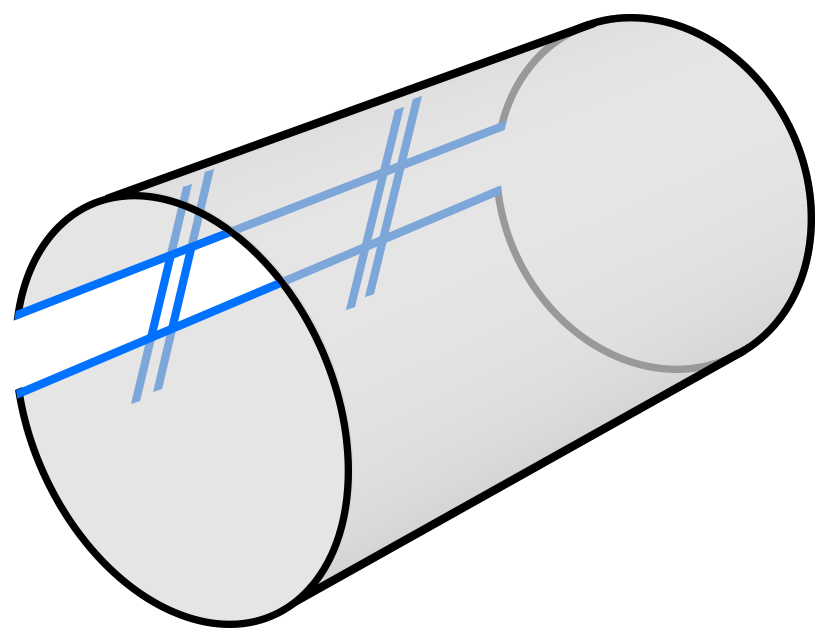}.
}
For a compact space dimension,  
\ali{
	|\psi\rangle = \adjincludegraphics[width=2.5cm,valign=c]{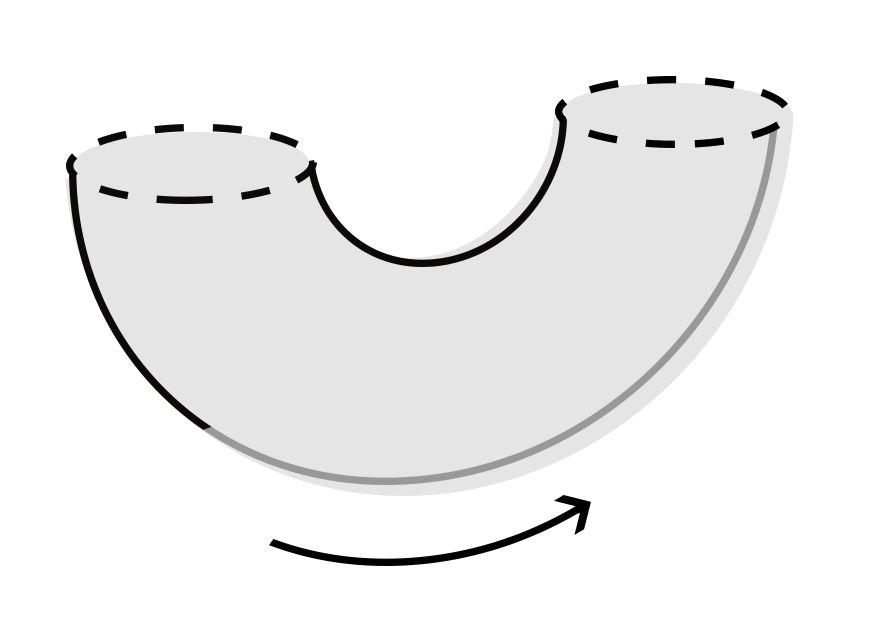}  \llap{
		\parbox[][1.2cm][b]{.7cm}{{\scriptsize$\tau$}}} \llap{
		\parbox[][2.cm][b]{1.35cm}{{\tiny $\beta/2$}}} \llap{
		\parbox[][-1.3cm][b]{2.2cm}{{\scriptsize$(\phi_L)$}}} \llap{
		\parbox[][-1.5cm][b]{0.8cm}{{\scriptsize$(\phi_R)$}}}   
	\label{TFDpicturecompact}
}
and 
\ali{ 
	\rho_L\equiv \tr_R\rho = \adjincludegraphics[width=2.5cm,valign=c]{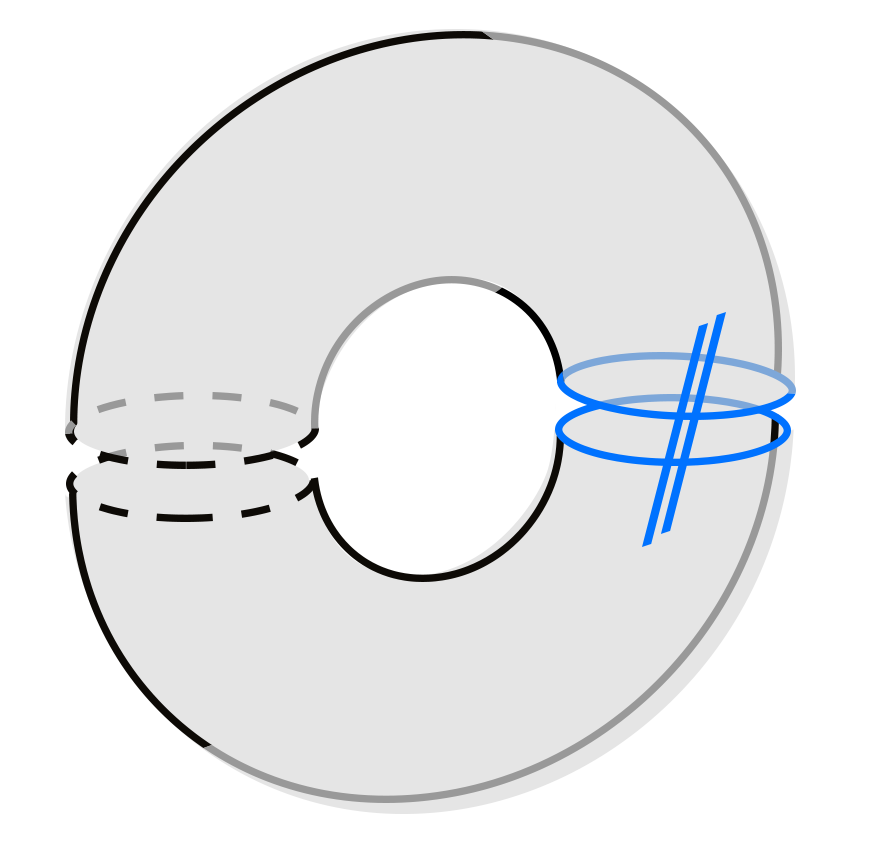} = \adjincludegraphics[width=2.5cm,valign=c]{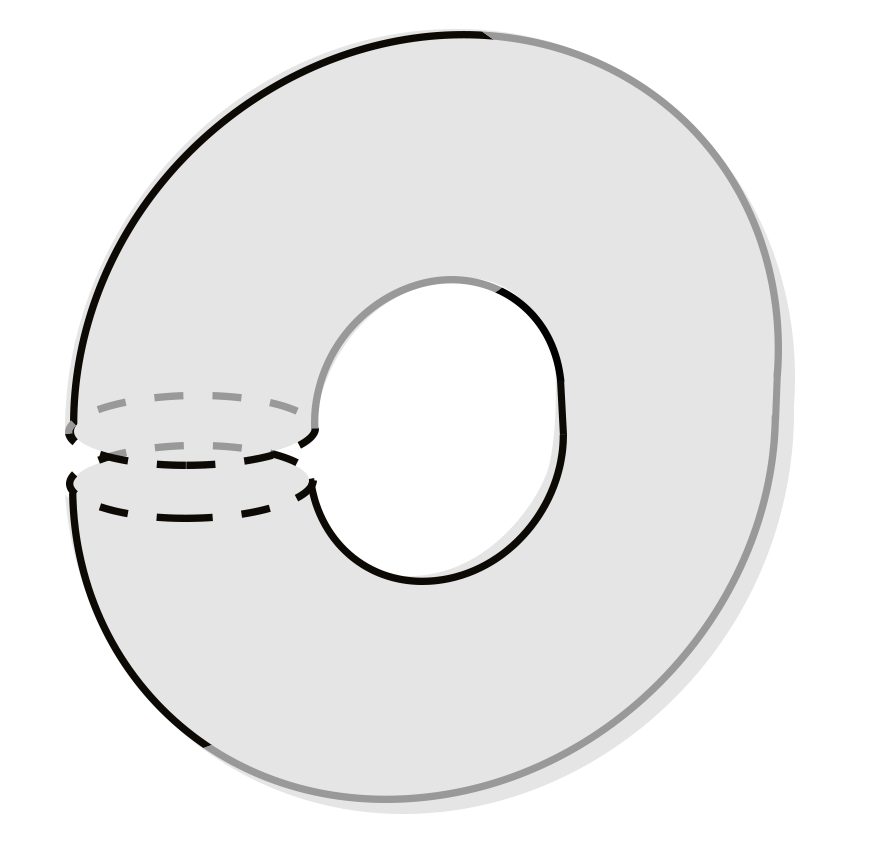} . 
}

The above statements are given in path integral language  
$|\psi \rangle = \int^\cdot \mathcal D \phi \, e^{-I[\phi]}$ 
or
\ali{
	\psi(\phi_L,\phi_R) = \int^{{\tiny{\begin{array}{ll} \phi(\tau_0)=\phi_L \\ \phi(\tau_0+\beta/2) = \phi_R \end{array}}}} \mathcal D \phi \, e^{-I[\phi]}, 
	\label{psiphiLphiR}
} 
as this gives an immediate visual derivation of \eqref{rhoATFD}-\eqref{TFDstatement1}. It is a simple exercise to derive \eqref{rhoATFD}-\eqref{TFDstatement1} from the explicit formula  
for the TFD state \eqref{TFDpicture} given by  
\ali{
	|\psi \rangle = \sum_n \sqrt{p_n} \,  |E_n\rangle_L |E_n\rangle_R, \qquad p_n = \frac{e^{-\beta E_n}}{Z(\beta)}  
}
with $|E_n\rangle$ the energy eigenstates of $H_L$ and $H_R$. 
It is a pure state per construction (using \eqref{rhoTFD}), and the construction is therefore referred to as purification. That is, $\tr(\rho^2) = \tr \rho$ and hence its von Neumann entropy is zero, while the von Neumann entropy of one copy $S_A = -\tr (\hat \rho_R \log \hat \rho_R)$ 
is the thermal entropy 
\ali{
	S_A = S_{thermal},  
}
where we used the notation $A$ for the region the observer of the `right' theory has access to. The TFD construction thus provides a set-up where the reduced density matrix is thermal \eqref{rhoATFD}, and the entanglement entropy $S_A$ is exactly given by a thermal entropy. 
As discussed in section \ref{sectionThermal}, this can be applied to the calculation of CFT entanglement of an interval $A$ by conformally mapping the interval to one copy of a TFD in Fig \ref{figHolzheypsi}. 

As is apparent from the path integral pictures above, a thermofield double  description arises for Euclidean manifolds that can be divided into two disconnected parts by the specification of \emph{two} values of a periodic coordinate $\tau$ (the values $\tau_0$ and $\tau_0 + \beta/2$ in \eqref{psiphiLphiR}), such as the cylinder or torus. 

\begin{figure}
\quad	\includegraphics[scale=0.18]{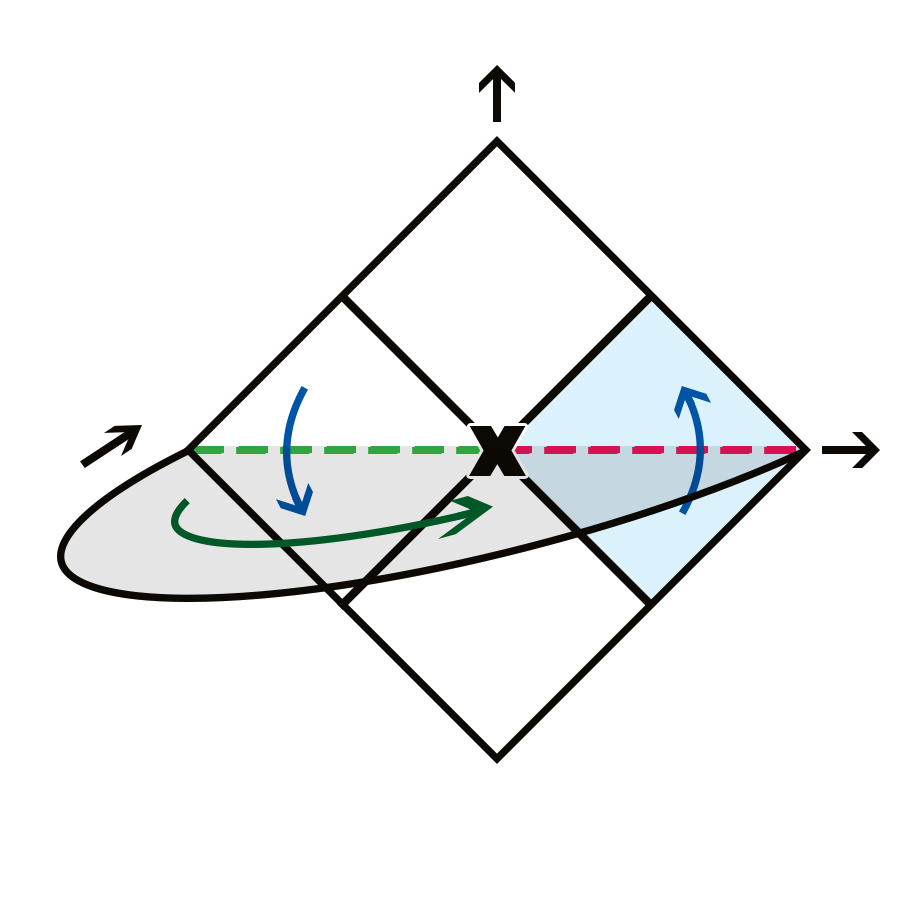} \llap{
		\parbox[][-7.6cm][b]{2.3cm}{$t$}} \llap{
		\parbox[][-3.6cm][b]{4.6cm}{$t_E$}} \llap{
		\parbox[][-4cm][b]{0.15cm}{$x$}}  \llap{
		\parbox[][-2.2cm][b]{3.5cm}{{\color{taugreen}$\tau$}}} \llap{
		\parbox[][-4.5cm][b]{3.1cm}{{\color{tSblue}$t_S$}}} \llap{
		\parbox[][-4.5cm][b]{1.84cm}{{\color{tSblue}$t_S$}}} 
	\includegraphics[scale=0.18]{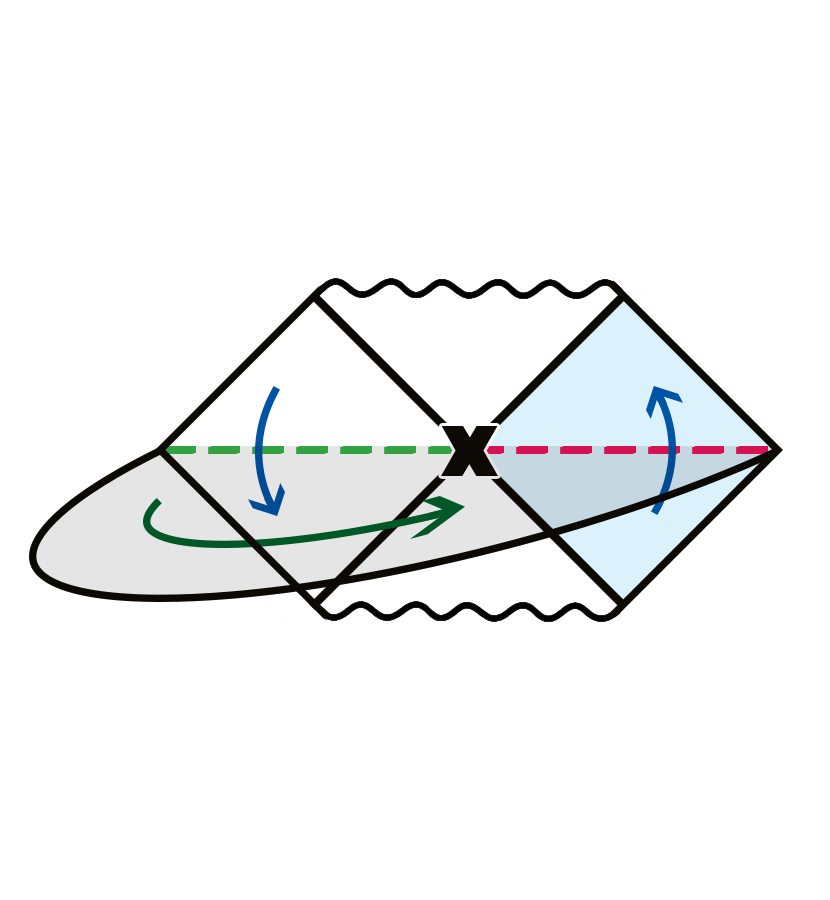} \quad	\includegraphics[scale=0.18]{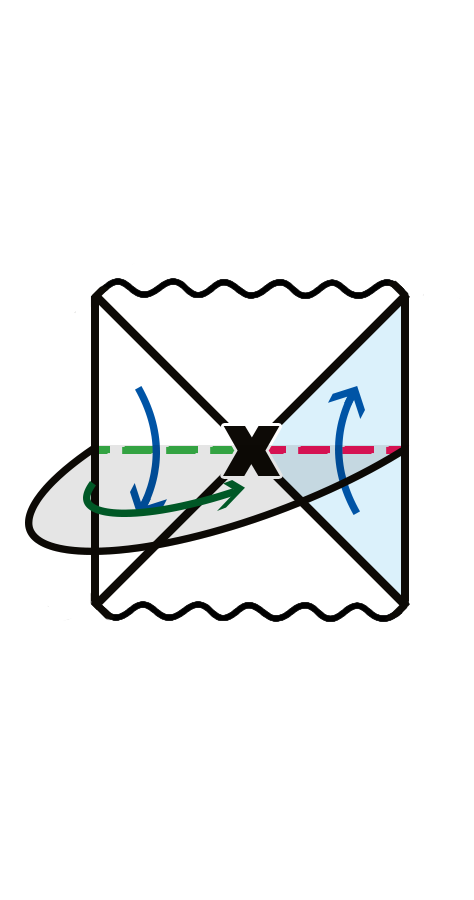}
	\caption{Penrose diagrams of a) Minkowski spacetime, b) the extended 
		Schwarzschild black hole, and c) the extended 
		AdS-Schwarzschild or BTZ black hole, covered by the full range of $(t,x)$ coordinates. In the case of the black holes, $(t,x)$ are the Kruskal coordinates. 
		Superimposed is the Euclidean preparation of the TFD state, which provides the Hartle-Hawking state.  
		The state can then be further evolved in Lorentzian time. 
		To an observer in resp.~the Rindler wedge, the (1-sided) Schwarzschild geometry and the (1-sided) BTZ geometry (all marked in light blue), the state appears thermal. 
		The (Lorentzian) time coordinate $t_S$ that covers these regions is Wick rotated, $t_S = i \tau$, to the periodic coordinate $\tau$ of the thermofield double construction.     
	}
	\label{figPenrose}
\end{figure}

Let us discuss two more examples. 
(For more details, see e.g.~\cite{Harlow:2014yka}.)     
The first is the Euclidean disk with metric $ds^2 = dR^2 + R^2 d\tau^2$. In polar coordinates with polar angle $\tau$, one can  
consider the TFD state 
\ali{
	|\psi\rangle = \adjincludegraphics[width=2.5cm,valign=c]{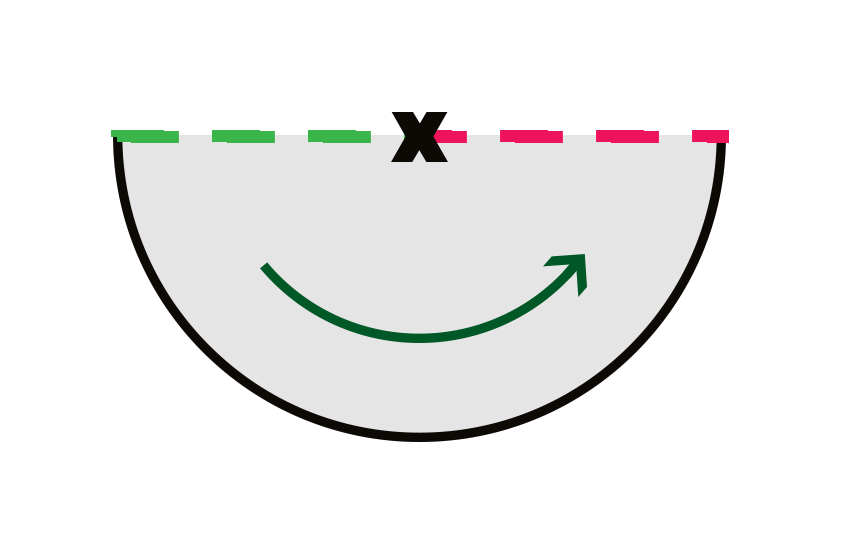}  \llap{
		\parbox[][-.2cm][b]{.7cm}{{{\color{taugreen}\scriptsize$\tau$}}}}
}
with 
\ali{
	\tr_L\rho = \adjincludegraphics[width=2cm,valign=c]{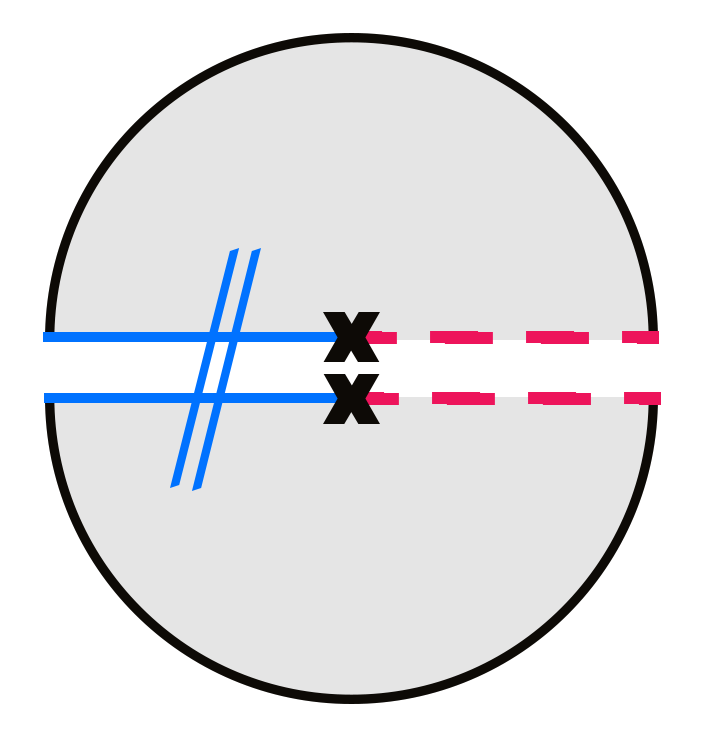} 
	= \adjincludegraphics[width=2cm,valign=c]{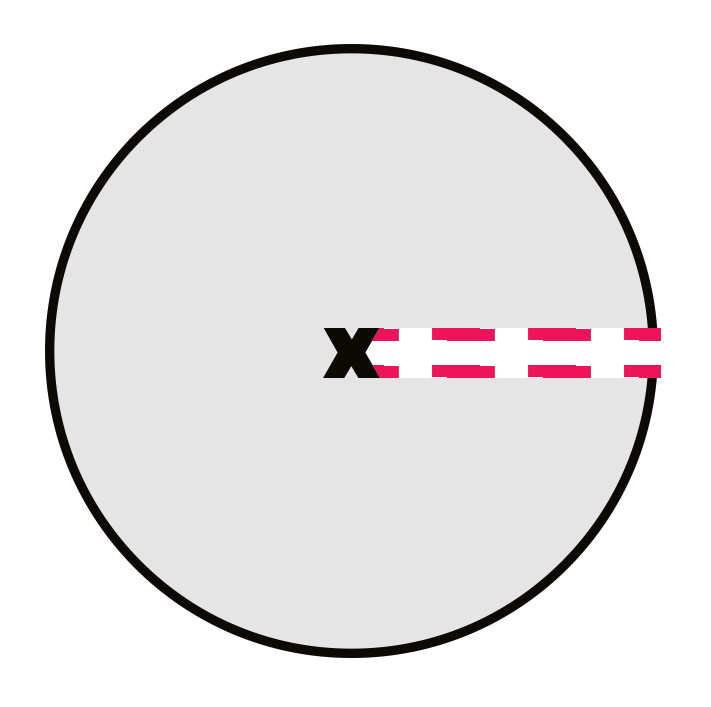} 
	= \rho_{thermal}   
}
the thermal density matrix encountered in Fig \ref{figHolzheyrhoA}b, for $\beta = 2\pi$ the inverse Unruh temperature \cite{Unruh:1976db} detected by a Rindler observer in Lorentzian signature (Fig \ref{figPenrose}a). 
Euclidean evolution prepares the vacuum state $|\psi\rangle$  
on the Minkowski plane $ds^2 = -dt^2 + dx^2$, at the spacelike slice located at $t=0=t_E$, i.e.~the intersection of zero Lorentzian and Euclidean time $t_E=i t$. 
A Rindler observer is a boosted observer (with acceleration set to one below) who only has access to the positive half-line $x > 0$ at $t=0$, with domain of dependence the Rindler wedge 
\ali{
	ds^2_{Rindler} &= -dt^2 + dx^2 , \qquad (x \geq 0, \, |t|<x) \\ 
	&= dR^2 - R^2 dt_S^2, \qquad (R\geq 0, \, \text{all } t_S) \label{dsRindler}
}
which is covered by Rindler coordinates $(R, t_S)$ or $(x_z, t_S)$ with $x_z = \log R$. These are related to Minkowski coordinates by $x = R \cosh t_S$, $t = R \sinh t_S$. The half-line $A$ ($x >0$) is effectively separated from the half-line $\bar A$ ($x<0$) by the Rindler horizon $R=0$.  

A last example is the cigar manifold, which interpolates between the Euclidean disk geometry (at the tip of the cigar) and the Euclidean cylinder. It appears in the Wick-rotated metric of black hole backgrounds, e.g.~the Schwarzschild black hole in Fig \ref{figPenrose}b. 
The associated TFD state $|\psi\rangle$ provides the Hartle-Hawking state \cite{Hartle:1976tp} of quantum fields on the black hole background, defined by doing the path integral over half the Euclidean geometry, 
and $\tr_L |\psi\rangle \langle \psi| = \rho_{thermal}$ describes a thermal state at the Hawking temperature of the black hole. The prepared state at the intersection of zero Euclidean and Lorentzian time can then be further  evolved in Lorentzian time $t$.

\paragraph{The holographic dual of the TFD} 

The extended $(2+1)$-dimensional AdS-Schwarzschild black hole or BTZ solution \cite{BTZ-Banados:1992wn} is shown in Fig \ref{figPenrose}c, with a Hartle-Hawking state prepared by evolution over the Poincar\'e disk $ds^2 = 4 dw d\bar w/(1 - w \bar w)^2$. 
This $(2+1)$-dimensional geometry has two asymptotic boundaries where two dual CFT copies live. It is the holographic dual of the TFD state of $(1+1)$-dimensional CFT \cite{Maldacena:2001kr}. Indeed, when the suppressed spacelike coordinate $x_z$ of the CFT is added back to the Euclidean half of the Penrose diagram in Fig \ref{figPenrose}c, the AdS TFD state at the conformal boundary becomes the CFT TFD state of Fig \eqref{TFDpicture} or \eqref{TFDpicturecompact}, depending on whether $x_z$ has an infinite or compact range.

\paragraph{An application of equation \eqref{Sconf}} 

\begin{figure}
    \centering
	\includegraphics[scale=0.18]{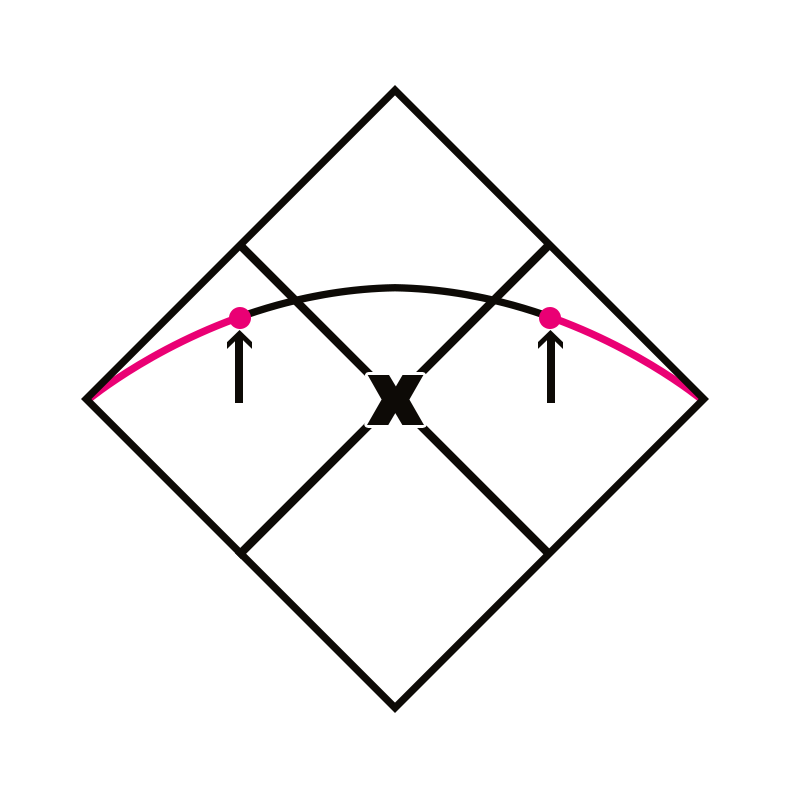} \llap{
		\parbox[][-3.6cm][b]{1.15cm}{$t_b$}} \llap{
		\parbox[][-3.6cm][b]{3.15cm}{$t_b$}}
	\caption{Region $A$ in red, with entanglement $S_A$ given in \eqref{SAlogcosh}. } 
	\label{figlogcosh}
\end{figure}

To end this section, we focus on the Minkowski geometry of Fig \ref{figPenrose}a and determine the Lorentzian time $t_b$-dependence of the entanglement of the region $A$ pictured in Fig \ref{figlogcosh}, following Hartman and Maldacena  \cite{Hartman:2013qma}. 
The region consists of the half-line $x_z>0$ in each Rindler wedge, in cylinder coordinates $dx_z^2 - dt_S^2$.

At time $t_b=0$, the Cauchy slice on which $A$ is defined is simply at $t=0$.  Then, point $P_1$ in the left Rindler wedge has Rindler time coordinate $\tau = 0$ and $x_z=0$, while point $P_2$ in the right Rindler wedge has coordinates $\tau = \beta/2$ and $x_z=0$ or $(z_2,\bar z_2) = (i \beta/2,-i \beta/2)$. Next, the Cauchy slice can be pushed upwards, with time coordinate $t_b$ equal to $t_S$ in the right Rindler wedge and equal to $-t_S$ in the left one. By analytic continuation, the locations of the interval endpoints $\p A$ are 
\ali{
	(z_1, \bar z_1) = (-t_b, t_b), \qquad (z_2, \bar z_2) = (t_b +  \frac{i\beta}{2}, -t_b - \frac{i\beta}{2}).  \label{HartmanMaldacenalocations}
} 
Applying equation \eqref{Sconf} with these values, and with $f(z) = \exp{(2\pi z/\beta)}$ the planar coordinate, one obtains 
\ali{
	S_A = \frac{c}{3} \log \left( \frac{\beta}{\pi \epsilon_z} \cosh(\frac{2\pi}{\beta} t_b) \right) \label{SAlogcosh}
}
for the time-dependence. At large times, the behavior is linear $S_A \sim t_b$. 

This is indeed $S_A = c/3  \log (\Delta x/\epsilon_{x})$ with the relation $x = e^{2\pi x_z/\beta} \cosh (2 \pi\, t_S /\beta)$ between planar and Rindler coordinates evaluated at $x_z=0$ at the interval endpoints, and 
$\epsilon_{x} = \frac{2\pi}{\beta} \epsilon_z$ by equation \eqref{cutoffs} with \eqref{HartmanMaldacenalocations}. 
The holographically dual calculation of $S_A$ is also performed in \cite{Hartman:2013qma}.

\chapter{Holographic entanglement} 

\abstract*{In this chapter we summarize the proof by Lewkowycz and Maldacena of the holographic interpretation of CFT entanglement as a minimal area, known as the `Ryu-Takayanagi' (RT) formula. This involves a bulk replica trick. We comment briefly on extensions of their proof that have been used to calculate the entropy of Hawking radiation with the so-called `island rule'. Then returning to holography, we discuss how the RT formula was initially proven for a special, U(1)-symmetric case by first mapping the CFT entanglement to a thermal entropy, for which a standard holographic dual interpretation is known. This discussion constitutes the holographic dual of the discussion on CFT entanglement as thermal entropy in the previous chapter. Finally, the intuition gained from the U(1) case proof of the RT prescription is used to discuss how gravitational first laws have interpretations in terms of CFT entanglement, leading to statements on the emergence of gravity from entanglement. We mention similar statements about emergent gravity from entanglement in non-holographic set-ups.}

Holography, in a general sense, posits there is a physically dual description of CFT in terms of a higher-dimensional theory of gravity. The CFT 
sets asymptotic boundary conditions for the gravitational theory, and the duality can be most succinctly stated as an equality of bulk (gravity) and boundary (CFT) partition functions 
\ali{
	Z_{grav} = Z_{CFT}.  \label{AdSCFT}
}
A subscript is included to refer to the partition function of the previous chapter $Z$ as the \emph{CFT} theory. No further knowledge of holography or AdS/CFT \cite{Maldacena:1997re} will be assumed. 

In this chapter we will discuss the holographic interpretation of the $\log L$ formula for the CFT entanglement \eqref{Sresult}. 
It famously has a geometric interpretation as a minimal area, 
according to the `Ryu-Takayanagi' (RT) formula \cite{Ryu:2006bv,Ryu:2006ef}.  
The proof of the RT formula by Lewkowycz and Maldacena \cite{Lewkowycz:2013nqa}, involving a bulk replica trick,  
is summarized in the first section.  It is henceforth referred to as the LM derivation. 
In recent years, extensions of their proof have been used to calculate the entropy of Hawking radiation with the so-called `island rule' and address the black hole information paradox. We comment on this briefly in section \ref{subsectislands}. 
Then returning to holography, we discuss in section \ref{sectholoholzhey} how the RT formula was initially proven for a special, U(1)-symmetric case by first mapping the CFT entanglement to a thermal entropy, for which a standard holographic dual interpretation is known. 
This discussion constitutes the holographic dual of the discussion on CFT entanglement as thermal entropy in section \ref{sectionThermal}. 
Finally, in section \ref{sectemergentgrav}, the intuition gained from the U(1) case proof of the RT prescription is used to discuss how gravitational first laws have interpretations in terms of CFT entanglement, leading to statements on the emergence of gravity from entanglement. 
Section \ref{subsectnonholo} comments briefly on similar statements about emergent gravity from entanglement in non-holographic set-ups.

\section[LM derivation of RT formula]{Lewkowycz Maldacena derivation of Ryu-Takayanagi formula} \label{sectionLM}

Here the set-up is a gravitational theory (e.g.~\cite{Gibbons:1978ac}) 
\ali{ 
	&Z_{grav} = \int \mathcal D g \mathcal D \phi e^{-I[g,\phi]}, \quad \\
	&I = -\frac{1}{16\pi G}\int d^{d+1} x \sqrt{g} (R - 2 \Lambda) + I_{GHY} + I_m[g,\phi]. 
}
The dynamics of the metric field $g$ with curvature $R$ is described by the Euclidean action $I$ of Einstein gravity, with negative cosmological constant $\Lambda$, 
Gibbons-Hawking-York boundary term $I_{GHY}$, and the action $I_m$ for the  matter fields $\phi$. We will focus, to begin with, on the $d=2$ case, interpreting the gravitational theory as the 3-dimensional, asymptotically Anti-de Sitter (AdS) dual of the 2-dimensional CFT of chapter \ref{chapterCFT}. By the AdS/CFT duality \eqref{AdSCFT}, the CFT entanglement $S_A = (1 - n \p_n) \log Z_{CFT}(n)|_{n \ra 1}$ discussed in the previous chapter has a dual interpretation as gravitational entropy
\ali{
	S_A = (1 - n \p_n) \log Z_{grav}(n)|_{n\ra 1}.   \label{gravreplica}
}
It is this object we are interested in calculating in this section, by employing a bulk version of the replica trick.   
 
The full, formal quantum gravity path integral $Z_{grav}$ has a semi-classical approximation $Z_{grav} \approx \exp{(-I[g_*,\phi_*])}$ in terms of the on-shell gravitational action, evaluated on a classical solution $(g_*,\phi_*)$. In this saddle point approximation, the gravitational entropy becomes calculable as   
\ali{
	S_A = -(1 - n \p_n) I[g_*,\phi_*](n)|_{n\ra 1} .    
} 
The solution has to satisfy boundary conditions  
set by the CFT replica manifold partition function $Z_{CFT}(n)$,   
with $g_*$ extending asymptotically to the manifold over which $Z_{CFT}(n)$ path integrates. From here on we will consider pure gravity in the bulk for simplicity, and at the end of the section mention results on extensions, such as the inclusion of matter fields $\phi$, 
higher derivative gravity, etc. 
 
Several incarnations of $Z_{CFT}(n)$ (with integer $n>1$) were discussed in the previous chapter: as the path integral of a replicated theory on the non-replicated orbifold manifold $\mathbb{C}$ (Fig \ref{figreplica}b), as the path integral of the non-replicated theory on the replicated manifold $\mathcal R_{n,A}$ (Fig \ref{figreplica}a) or as the path integral of the non-replicated theory on the replicated cylinder (Fig \ref{figZncylCFT}b). The last one provides the right starting point for constructing the semi-classical $Z_{grav}(n)$, because it is a smooth replica manifold, with a smooth corresponding bulk solution. 
A bulk geometry that extends the conical singularity of $\mathcal R_{n,A}$ into the bulk is not an acceptable solution of the sourceless Einstein equations that allows for a saddle point evaluation 
\cite{Headrick:2010zt}\footnote{
	An interpretation of the bulk configurations considered in \cite{Fursaev:2006ih} is discussed in \cite{Dong:2018seb}. 
}. 
Boundary conditions are imposed at the conformal boundary of the asymptotically AdS theory, thus allowing to use the replicated cylinder, which is conformally related to $\mathcal R_{n,A}$.

\begin{figure}
	\centering  
	 \includegraphics[width=3cm]{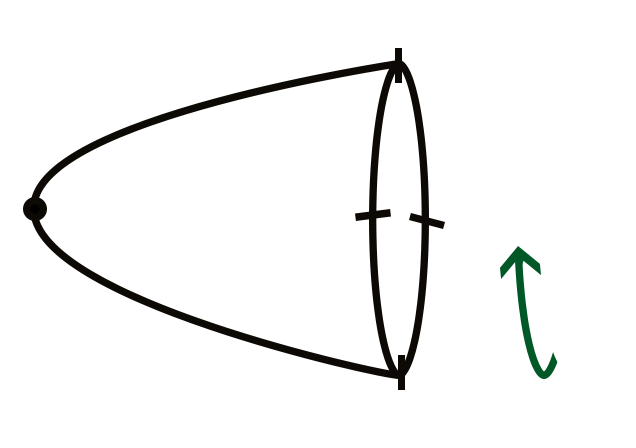} \llap{
		\parbox[][-1.5cm][b]{0.45cm}{{\color{taugreen} $\tau$}}}\llap{
		\parbox[][-1.7cm][b]{3.25cm}{$\Sigma$}} \qquad  \qquad  \includegraphics[width=3cm]{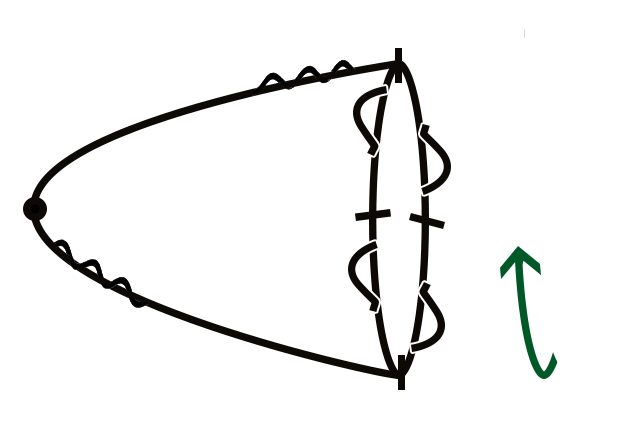} \llap{
		\parbox[][-1.5cm][b]{0.45cm}{{\color{taugreen} $\tau$}}} \llap{
		\parbox[][-1.7cm][b]{3.3cm}{$\Sigma$}}
	\caption{a) Bulk replica manifold 
		$\mathcal M_n$ (here pictured for $n=4$) that on the conformal boundary $\p \mathcal M_n$ reduces to the cylinder replica manifold of Fig \ref{figZncylCFT}b (with the spacelike direction of the CFT suppressed), for the density matrix $\rho_A = \exp{(-\beta H_\tau)}$ of \eqref{BWtheorem}. The uncontractible circle in the boundary becomes contractible in the bulk at $\Sigma$.    
		b) Bulk replica manifold $\mathcal M_n$ for more general, non-$U(1)$ symmetric case $\rho_A = \mathcal P \exp{\int_{\beta}^{0} d\tau \mathcal H(\tau)}$. 
	}
	\label{figLMsol}
\end{figure}

The replica cylinder of Fig \ref{figZncylCFT}b extends into the bulk in a (by ansatz) $\mathbb{Z}_n$ symmetric solution $\mathcal M_n$ that smoothly ends in a $\mathbb{Z}_n$ fixed point, as illustrated in Fig \ref{figLMsol}a. 
With the boundary replica manifold describing a thermal state, $\tr \rho_A^n = \tr \exp{(-n \beta H_\tau)}$, $\mathcal M_n$ is a black hole solution and the bulk replica method will reduce to the Gibbons-Hawking path integral derivation of its Bekenstein-Hawking entropy (see section \ref{sectholoholzhey}), providing the dual of the CFT entanglement $S_A$ for 
$\tau$ an angular coordinate going around the boundary $\p A$ of the interval $A$, as in Fig \ref{figHolzheyrhoA}. 
The relevant conformal mapping employed there  
\ali{
	ds^2 = R^2 d\tau^2 + dR^2 
	\, \ra \, d\tau^2 + \frac{dR^2}{R^2}
}
maps the $\tau$-direction on the plane to the uncontractible circle $S^1$ 
of a hyperbolic cylinder $S^1(\beta) \times \mathbb{H}_1$ 
(easily extended to $S^1(\beta) \times \mathbb{H}_{d-1}$ in the $d>2$ case, with corresponding replica cylinder $S^1(n\beta) \times \mathbb{H}_{d-1}$).  The hyperbolic cylinder has a U(1) isometry generated by the Killing vector $\p_\tau$, reflecting the conservation of the Hamiltonian $H_\tau$. 
However, the present discussion crucially is more generally valid: one can 
take as boundary conditions a CFT state 
\ali{
	\rho_A = \mathcal P e^{-\int_{0}^{\beta} d\tau \mathcal H(\tau)} 
	}
with Euclidean Hamiltonian density $\mathcal H(\tau)$ that is not necessarily U(1) invariant, but instead in general depends on $\tau$. 
The more general set-up is the following. Consider a spatial region $A$ in a $d$-dimensional boundary CFT at time equal to zero. 
Going to Euclidean signature, one can \emph{locally} at the boundary $\p A$ of $A$ choose an angular coordinate $\tau$ that has the property that it encircles $\p A$. For this coordinate, locally the mapping to a hyperbolic cylinder will be valid, but globally $\tau$-dependence can appear in the conformally scaled boundary metric, i.e.~U(1) invariance is lost. Still, the $\tau$-dependence has to be consistent with the $2\pi$-periodicity of the circle that $\tau$ parametrizes in the original geometry. That is, the (conformal) boundary metric can depend on $\tau$ through powers of $\exp{(\pm i \tau)}$. We follow \cite{Lewkowycz:2013nqa} in setting $\beta = 2\pi$.

The more general set-up is still characterized by a non-contractible $\tau$-circle $S^1(2\pi)$ in the boundary metric. The corresponding replica manifold is obtained by gluing together $n$ circles to $S^1(2\pi n)$, and is thus $\mathbb Z_n$ symmetric by construction, as a consequence of the periodicity in $\tau$. 
The bulk replica manifold $\mathcal M_n$ is then constructed as the (smooth) solution of the sourceless Einstein's equations that extends the replica manifold into the bulk in a $\mathbb Z_n$ symmetric way. Its general form, with $\mathbb Z_n$ fixed point $\Sigma$ where the boundary $\tau$-circle $S^1(2\pi n)$ becomes contractible, is illustrated in Fig \ref{figLMsol}b. 
To be more precise, $\Sigma$ is a set of points of co-dimension 2 in the $(d+1)$-dimensional bulk.  

Now we are ready to evaluate the entropy 
\ali{
	S_A = -(1 - n \p_n) I[\mathcal M_n]|_{n\ra 1} .    
}
This can be written as $S_A = n \p_n \left( I[\mathcal M_n] - n I[\mathcal M_1] \right)|_{n\ra 1}$ in terms of two smooth manifolds, $\mathcal M_n$ and $\mathcal M_1$, with different boundaries $\p \mathcal M_n \neq \p \mathcal M_1$. In order to compare them, they should be brought in a form where each  
has the same boundary. 
To achieve this, periodicity in $\tau$ (or said otherwise, $\mathbb Z_n$ symmetry) can be used to rewrite $I[\mathcal M_n]$ in the first term\footnote{
	Alternatively, one can proceed by rewriting $n I[\mathcal M_1]$ in the second term, into an action evaluated on a manifold with boundary $\p \mathcal M_n$ and conical excess $2\pi(n-1)$ \cite{Lewkowycz:2013nqa,Camps:2013zua}. After adding and subtracting the action evaluated on the manifold with the same boundary but a regulated, rounded off conical singularity, and using the fact that $\mathcal M_n$ is a solution of the equations of motion, the remaining contribution to $S_A$ comes from the regulated conical tip and is  proportional to its area \cite{Fursaev:1995ef,Fursaev:2013fta}. 
	\label{Campsfootnote}
} 
\ali{
	I[\mathcal M_n] = \int_0^{2\pi n} \mathcal L[g(n)] d\tau 
}
as an integer $n$ times an auxiliary action $I[\mathcal O_n]_{\text{smooth}}$ defined as 
\ali{
	I[\mathcal O_n]_{\text{smooth}} \equiv \int_0^{2\pi} \mathcal L[g(n)] d\tau. 
}
Here, $\mathcal L$ is the Lagrangian density after integrating over all but the $\tau$ coordinate, and $\mathcal O_n$ the orbifold manifold $\mathcal O_n \equiv \mathcal M_n/\mathbb Z_n$ obtained from the replica manifold by quotienting the replica symmetry, with boundary $\p \mathcal O_n = \p \mathcal M_1$.    
The relation 
\ali{
	I[\mathcal M_n] = n \, I[\mathcal O_n]_{\text{smooth}}  \label{LMtrick}
}
is well-defined also for non-integer $n$, and it will be used below to obtain $S_A$ in the $n \ra 1$ limit, or $\epsilon \ra 0$ for $n = 1 + \epsilon$. 

Let us discuss the orbifold for different values of $n$. 
For $n=1$, it is equal to the original dual of the CFT, $\mathcal O_1 = \mathcal M_1$, which we know is a solution of the $I[g]$ equations of motion.   
As it extremizes the action, this implies   
\ali{
	I[\mathcal O_{1+\epsilon}] = I[\mathcal O_1].  \label{LMonshell}
}
For $n \neq 1$, the orbifold has a conical singularity at the location of the $\mathbb Z_n$ fixed point in $\mathcal M_n$, with a deficit angle of $2\pi(n-1)/n$ or opening angle $2\pi/n$. 
The corresponding curvature singularity, which is proportional to the deficit angle, contributes a term $I_{\text{sing}} \sim \int_{\text{sing}} \sqrt{g} R $ to the gravitational action evaluated on $\mathcal O_n$,  
\ali{
	I[\mathcal O_n] = I[\mathcal O_n]_{\text{smooth}} + I_{\text{sing}}. \label{IOsmooth}  
}
The first term contains the regular contribution $I[\mathcal O_n]_{\text{smooth}} \sim \int_{\text{cone} \setminus \text{sing}} \sqrt{g} R $. 
It is this regulated contribution that appears in \eqref{LMtrick} as a rewriting of the \emph{smooth} left hand side.  

Because of the conical singularity, $\mathcal O_n$ is not a solution of the $I[g]$ equations of motion, but of Einstein's equations in the presence of a source localized at the singularity. The required source is a co-dimension 2 brane (in the $(d+1)$-dimensional bulk) of tension $T = (n-1)/(4 n G)$ and action $I_{\text{brane}} = T \int d^{d-1} y \sqrt{h}$, for $y$ and $h$ the induced coordinates and metric on the brane \cite{Dong:2016fnf}.  
This means that on-shell, 
\ali{
	I[\mathcal O_n] + I_{\text{brane}} &= I[\mathcal O_n]_{\text{smooth}} \label{LMbrane} 
}
or said otherwise, that $I_{\text{sing}} = -I_{\text{brane}}$, which can be checked explicitly. 	In evaluating $I[\mathcal O_n]_{\text{smooth}}$ one can thus think of it either as integrating the gravitational Lagrangian over the regulated cone with no contribution of the singularity \eqref{IOsmooth}, or alternatively as \emph{adding} the brane \eqref{LMbrane}: $\int_{\text{cone}\setminus \text{sing}} \mathcal L d\tau dr d\vec x  = \int_{\text{cone}} \mathcal L d\tau dr d\vec x + \int_{\Sigma} \mathcal L_{brane} d\vec x$ (with here $\mathcal L$ the gravitational Lagrangian density). 

Next we can write $I[\mathcal M_{1+\epsilon}]$ in terms of $I[\mathcal O_n]_{\text{smooth}}$ by \eqref{LMtrick}, and then use \eqref{LMbrane} and finally \eqref{LMonshell} to obtain 
\ali{
	I[\mathcal M_{1+\epsilon}] = (1 + \epsilon) I[\mathcal M_{1}] + \epsilon \frac{\mathcal A[\Sigma_{min}]}{4G}. 
} 
The last term is $I_{\text{brane}}$ in the limit of a tensionless brane $\epsilon \ra 0$, which does not backreact and settles in the location $\Sigma$ that minimizes its Nambu-Goto `worldvolume' or area $\mathcal A[\Sigma_{min}]$. The entropy $S_A$, with the limit reexpressed for $n = 1+\epsilon$, 
\ali{ 
	S_A = \left. \frac{I[\mathcal M_{1+\epsilon}] - I[\mathcal M_1]}{\epsilon} - I[\mathcal M_1] \,\, \right|_{\epsilon \ra 0}    
}
finally becomes 
\ali{
	S_A = \frac{\mathcal A[\Sigma_{min}]}{4G}. \label{LMresult}
	}
This is the Ryu-Takayanagi (RT) formula for the holographic entanglement of a spacelike region $A$. 
It also includes a homology condition, which has been discussed in the context of the LM derivation in \cite{Haehl:2014zoa}. 

For general time-dependent states, the area should be extremized according to the HRT formula \cite{Hubeny:2007xt}. A derivation of this covariant prescription was given in \cite{Dong:2016hjy}.  
The generalization to higher derivative gravity involving Wald entropy \cite{Iyer:1994ys,Wald:1993nt} was derived in \cite{Dong:2017xht}. 

The inclusion of matter fields -- in a semi-classical treatment of the bulk theory -- lead to a first-order correction $\mathcal O (G^0)$ to the extremal area of bulk matter entropy $S_m$ \cite{Lewkowycz:2013nqa,FLM-Faulkner:2013ana}.  
It is  consistent at that order with the later proposed  
Engelhardt-Wall formula \cite{Engelhardt:2014gca}, at arbitrary orders in the bulk Planck constant,   
\ali{
	S_A = S_{gen}[\Sigma_{QES}] \label{QES}
} 
where the `quantum extremal surface' $\Sigma_{QES}$ is the minimal co-dimension 2 surface that extremizes the bulk generalized entropy $S_{gen} \equiv \mathcal A/4G + S_m(EW)$. Here, $EW$ is short for entanglement wedge, defined as the bulk region confined by $A$ and its RT surface $\Sigma$, and $S_m(EW)$ is the von Neumann entropy of bulk matter fields in $EW$.   
The QES prescription was derived in \cite{Dong:2017xht} and in \cite{Almheiri:2019qdq} by inserting twist fields at $\Sigma$ for the replicated matter sector in the orbifold $\mathcal O_n$, see also \cite{Pedraza:2021ssc}.

\subsection{Islands}  \label{subsectislands}

The gravitational entropy formula \eqref{QES} has also been applied beyond the context of holography in set-ups that exhibit (a version of) a black hole information paradox to calculate the entropy of Hawking radiation. Without going into detail, we will sketch the idea. 

\begin{figure}
	\centering \includegraphics[scale=0.18]{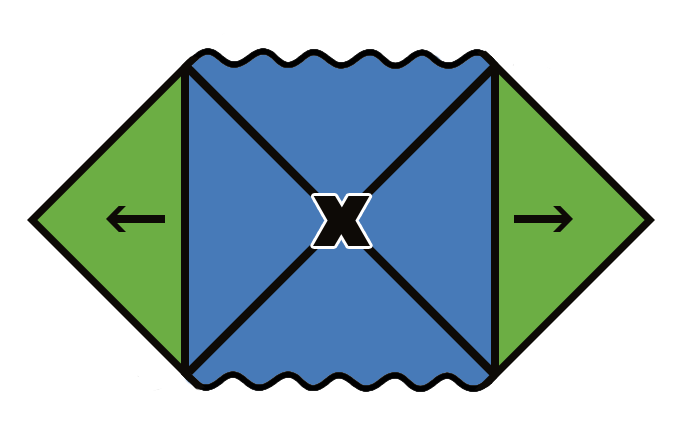}\llap{
		\parbox[][-2.1cm][b]{2.75cm}{$x_z$}} \llap{
		\parbox[][-2.1cm][b]{.85cm}{$x_z$}} \qquad \qquad \qquad 
	\includegraphics[scale=0.18]{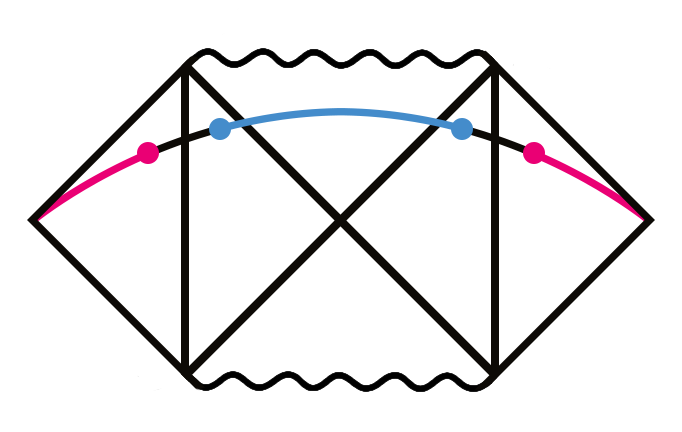} \llap{
		\parbox[][-2.4cm][b]{1.75cm}{{\color{islandblue} $I$}}} \llap{
		\parbox[][-1.65cm][b]{0.75cm}{{\color{radiationred} $R$}}} \llap{
		\parbox[][-1.65cm][b]{3.1cm}{{\color{radiationred} $R$}}}
	\caption{Left: The set-up for the calculation of radiation entropy, in Lorentzian signature. It consists of an AdS$_2$ black hole (blue) with a bath region that collects black hole radiation attached (green). For visualization one can think of the bath regions as extending backwards. Right: Radiation region $R$ with entropy given by the formula \eqref{islandSgen}, which contains contributions from a possible island region $I$. 
} \label{figIsland}
\end{figure}
\begin{figure}
	\centering \includegraphics[scale=0.17]{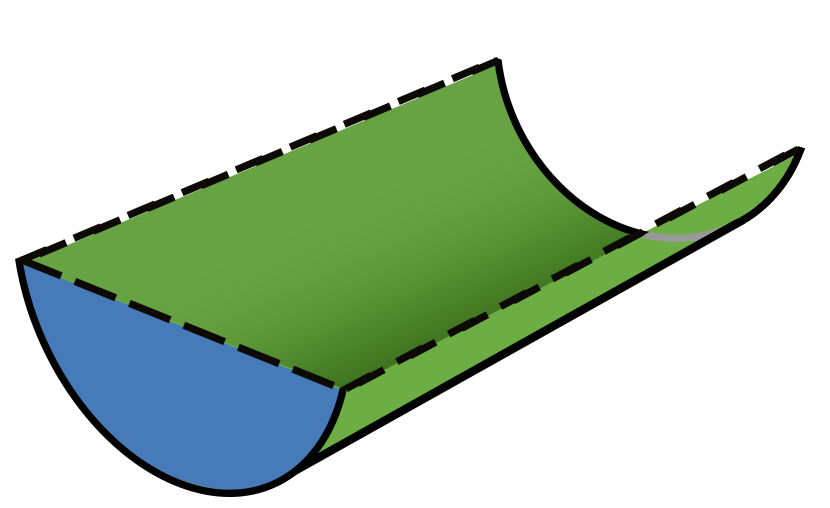}  
	\qquad \qquad \qquad
	\includegraphics[scale=0.17]{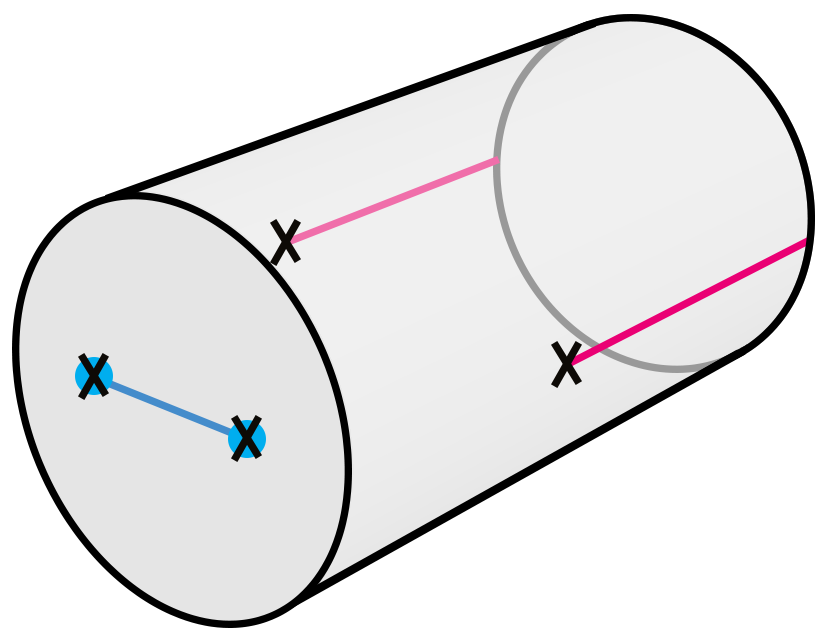} \llap{
		\parbox[][-2.1cm][b]{3.05cm}{{\color{islandblue} $I$}}} \llap{
		\parbox[][-3.2cm][b]{0.75cm}{{\color{radiationred} $R$}}} \llap{
		\parbox[][-4.05cm][b]{2.2cm}{{\color{radiationred} $R$}}}
	\caption{Left: 
		The state is the combined TFD state of the black hole and the bath region, consisting of the Poincar\'e disk evolution of Fig \ref{figPenrose}c in blue and $x_z>0$ cylinder evolution of \eqref{TFDpicture} in green.  
		Right: Orbifold with twist fields (crosses) at boundary $\p R$ of radiation region $R$ (at $t_b=0$)  
		and twist fields as well as branes (blue dots) at boundary $\p I$ of possible island region $I$.    
	}
	\label{figIsland2}
\end{figure}

The set-up considered in \cite{Almheiri:2019qdq} is given by a 2-dimensional AdS-black hole solution of Jackiw-Teitelboim gravity \cite{Jackiw:1984je,Teitelboim:1983ux,Almheiri:2014cka}  
connected to a non-gravitational bath region that collects black hole radiation. There is a matter CFT of large central charge $c \gg 1$  
that extends over both the gravitational and non-gravitational regions.  
The bath consists of an $x_z>0$ Minkowski geometry attached to the right asymptotic boundary of AdS and another copy to the left (Fig \ref{figIsland}a). 
One can prepare the state at time zero by Euclidean evolution over half the Euclidean AdS$_2$-black hole solution and half the $x_z>0$ cylinder, resulting in the combined TFD state of the black hole and the bath region (Fig \ref{figIsland2}a).  
The set-up is entirely 2-dimensional and non-holographic, but does have a holographic interpretation \cite{Almheiri:1908hni,Chen:2020uac} where the AdS$_2$ region forms an end of the world brane of a 3-dimensional AdS bulk.

Next, one can consider a `radiation' region $R$ extending from spacelike infinity to a value of $x_z$ greater than zero in each copy of the bath. The entanglement $S_R$ of CFT fields in $R$ can be calculated by the replica trick of section \ref{sectionreplicaCFT}. 
Step one is to construct the replica manifold consisting of $n$ copies of the system with cuts along $R$ that are glued cyclically, as in Fig \ref{figreplica}a. Upon analytic continuation, the dependence of $S_R$ on time $t_b$ going forward in each copy of the theory can be obtained, as in the derivation of equation \eqref{SAlogcosh}. The result is a continuously rising $S_R(t_b) = \frac{c}{3} \log(2 \cosh t_b)$, 
the Hawking entropy for black hole radiation collected in $R$. The absence of a Page curve for this entropy constitutes a black hole information paradox \cite{Almheiri:1910yqk}.  

The gravitational region in the replica manifold is left to dynamically determine its geometry, given the boundary conditions fixed at its edge. 
The replica manifold giving rise to the Hawking entropy only has connections between the $n$ sheets along $R$, and disconnected gravitational regions. 
Now the insight of \cite{Almheiri:2019qdq} 
was that connections between the $n$ sheets in the gravitational region also  
can arise, 
when a gravitational solution is taken into account that represents a (replica symmetric) wormhole between the replicas\footnote{
	For the role of wormholes in replica trick 
	computations, see also \cite{Penington:2019kki},  
	reviewed in \cite{Mertens:2022irh}. 
}. The connection between the sheets within the gravitational region is along a region $I$ called the island. In the orbifold $\mathcal O_n = \mathcal M_n/\mathbb Z_n$ of the replica manifold $\mathcal M_n$, pictured in Fig \ref{figIsland2}b, the endpoints $\p R$ of the radiation region are smooth and carry twist fields (as in Fig \ref{figreplica}b), while the endpoints $\p I$ of the island region have conical deficits (as in \eqref{IOsmooth}-\eqref{LMbrane}) and carry both twist fields and branes. An extension of the Lewkowycz Maldacena argument in section \ref{sectionLM} leads to the QES formula \eqref{QES} for the entropy $S_R$ as the  minimal generalized entropy \cite{Almheiri:2019qdq} 
\ali{ 
	S_{gen} = \frac{\Phi(\p I)}{4G} + S_m(R\cup I),  \label{islandSgen}   
} 
with $\Phi$ the Jackiw-Teitelboim dilaton and $\p I$ the location extremizing $S_{gen}$.   
The replica manifold with the island produces a radiation entropy expression $S_R$ that becomes minimal for large $t_b$, giving rise to the sought-after Page curve.

\section{Holographic entanglement and thermal entropy} \label{sectholoholzhey} 

In this section, we focus on the U(1) symmetric case of the holographic entanglement of section \ref{sectionLM}, or the Casini Huerta Myers derivation \cite{Casini:2011kv} of the Ryu-Takayanagi formula for the CFT entanglement \eqref{Sresult} of an interval. This will provide useful intuition for the discussion of the entanglement first law and its gravitational dual in the next section. 
  
The simplification in the U(1) case is that the entanglement maps to a thermal entropy, and one can construct the direct holographic dual interpretation of the CFT discussion of section \ref{sectionThermal}. This will lead to Fig \ref{figHolzheyholo} presenting the bulk equivalent of the mapping pictured in Fig \ref{figHolzheypsi}, and to the RT formula coinciding with the black hole entropy formula.

\subsection{Holographic thermal entropy}

\begin{figure}
	\centering \includegraphics[width=4cm]{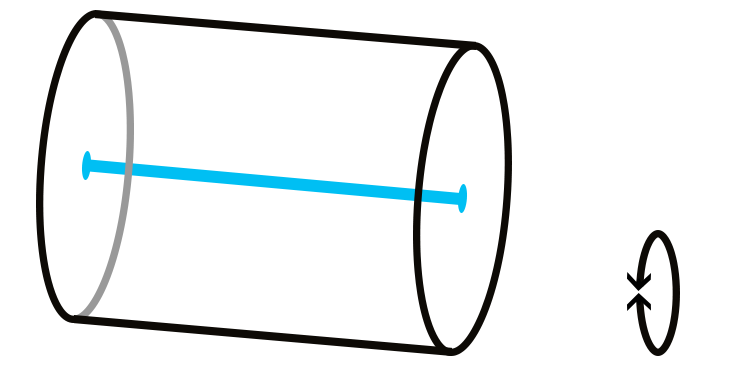} \llap{
		\parbox[b]{4.5cm}{a)\\\rule{0ex}{2.2cm}}} \llap{
		\parbox[][-1.5cm][b]{1.57cm}{$r=r_+$}} \llap{
		\parbox[][-0.5cm][b]{1.1cm}{$\beta$}}
	\qquad \qquad   \includegraphics[width=2cm]{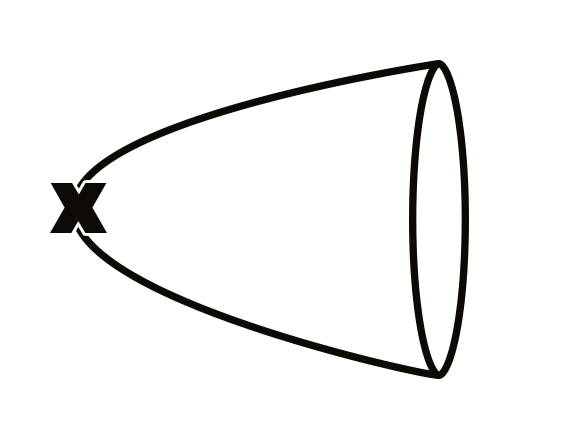} \llap{
		\parbox[b]{2.9cm}{b)\\\rule{0ex}{2.2cm}}} \llap{
		\parbox[][-1.3cm][b]{2.35cm}{$r_+$}} \quad \includegraphics[width=2cm]{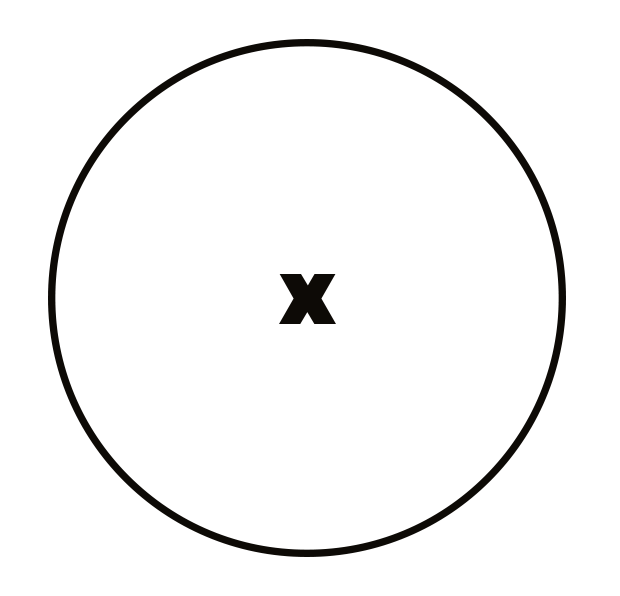} \llap{
		\parbox[][-1.5cm][b]{1.49cm}{$r_+$}}
	\caption{a) The thermal CFT partition function of Fig \ref{figZncylCFT}a is equated, by AdS/CFT, to the thermal partition function $Z_{grav}(\beta)$ of a black string solution of AdS gravity, represented by the corresponding solid  cylinder. b) An $(r,\tau)$ slice of the solid cylinder pictured as a cigar, with U(1) fixed point at $r = r_+$, or a Poincar\'e disk $ds^2 = 4 dw d\bar w/(1 - w \bar w)^2$ with $w = \sqrt{(r-r_+)/(r+r_+)} \exp{(i r_+ \tau)}$.   	 
	}
	\label{figBTZ}
\end{figure}

Let us first 
consider the $(2+1)$-dimensional BTZ black string solution 
\ali{
	ds^2 = -(r^2 - r_+^2) dt_S^2 + \frac{dr^2}{r^2-r_+^2} + r^2 dx_z^2  \label{BTZstring}
}
with horizon at $r=r_+$, AdS boundary at $r \ra \infty$ and AdS radius $l=1$. The ranges of the coordinates are $t_S = -\infty..\infty, r\geq r_+$, and $x_z=-\infty..\infty$. The conformal boundary in Euclidean signature ($t_S = i \tau$) is the Euclidean cylinder $ds^2_{CFT} = d\tau^2 + dx_z^2$, with Euclidean time periodicially identified $\tau \sim \tau + \beta$. 
The uncontractible $\tau$-circle in the dual CFT becomes contractible in the bulk at $r=r_+$ (where $(r^2 - r_+^2) d\tau^2 \ra 0$). This typical feature of a solution with a horizon is represented in Fig \ref{figBTZ}b by a cigar geometry in the $(r,\tau)$ directions. Smoothness of the geometry near the tip of the cigar requires the periodicity of $\tau$ to be fixed in terms of $r_+$ to (e.g.~\cite{Cadoni:2010ztg})  
\ali{
	\beta = \frac{2\pi}{r_+}. \label{betarp} 
}   
This is the inverse Hawking temperature of the black string, and the Euclidean gravitational path integral $Z_{grav}(\beta)$, shown in Fig \ref{figBTZ} as the solid cylinder, is interpreted as a thermal partition function $\tr \rho_{grav}$.  
The corresponding entropy 
\ali{
	S_{thermal} = (1 - \beta \p_\beta) \log Z_{grav}(\beta) \label{replicagrav}
}
is the holographic dual of the thermal entropy \eqref{eqSthermal} of the CFT. 
It can be calculated in different ways, as is nicely summarized in  e.g.~\cite{Balasubramanian:2013rqa}. The original Gibbons-Hawking method \cite{Gibbons:1976ue} took an `on-shell approach': as $\beta$ is varied, so is the mass (and thus the horizon radius $r_+$) of the black hole, in such a way that the on-shell relation \eqref{betarp} is maintained. 
Variation of the on-shell action $(-\log Z_{grav}(\beta))$ \cite{Hartmanlectures,Balasubramanian:1999re} 
then gives rise to the famous Bekenstein-Hawking entropy of the black hole  \ali{
	S_{thermal} = \frac{\mathcal A_h}{4G},  \label{BHentropy}
} 
where $\mathcal A_h$ is the horizon area. 
Another option is the `off-shell approach': when only $\beta$ is varied, the tip of the cigar turns into a conical singularity. As discussed at length in section \ref{sectionLM}, such a singular geometry does not solve the (sourceless) gravitational equations of motion. 
\eqref{replicagrav} can then be written in terms of the conical excess $\epsilon$, with $\delta \beta/\beta = \delta \epsilon$ \cite{Balasubramanian:2013rqa,Nelson:1994na}, as 
\ali{
	S_{thermal} = (1 - \p_\epsilon) \log \tr \rho_{grav}^{1+\epsilon} |_{\epsilon \ra 0} = (1 - n \p_n) \log \tr \rho_{grav}^{n} |_{n \ra 1}  , 
	\label{Sthermalreplica}
}
reproducing for $n = 1+\epsilon$ the replica trick formula \eqref{gravreplica}. 
Indeed the Gibbons-Hawking entropy derivation in this approach reduces to a special case of the gravitational replica trick\footnote{
	It is more directly related to the Lewkowycz Maldacena derivation following the option in footnote \ref{Campsfootnote}.  
} of section \ref{sectionLM}: the result \eqref{LMresult} for the entropy reduces to \eqref{BHentropy}.   
It is a special case because the gravitational solution \eqref{BTZstring} has a U(1) Killing vector $\p_{t_S}$, describing a system in thermal equilibrium.

\subsection{Ryu-Takayanagi from thermal entropy argument}

\begin{figure}
	\centering \includegraphics[scale=0.17]{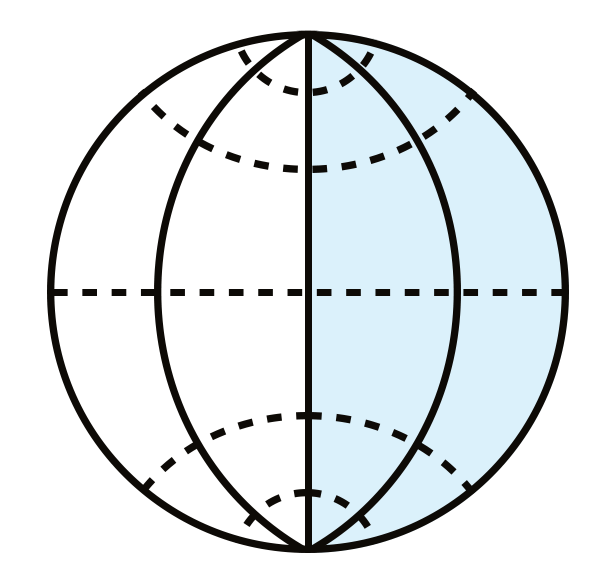} \llap{
		\parbox[][-1.8cm][b]{1.6cm}{\rotatebox{90}{$r=r_+$}}} \llap{
		\parbox[][0.5cm][b]{1.5cm}{$x_z=-\infty$}} \llap{
		\parbox[][-5cm][b]{1.5cm}{$x_z=\infty$}}
	\caption{A Poincar\'e disk constant time slice of AdS$_3$ in hyperbolic coordinates, with lines of constant $r$ (solid) and constant $x_z$ (dashed). The BTZ string metric \eqref{BTZstring} covers one half of the disk.} 
	\label{figBTZstringcoord}
\end{figure} 

\begin{figure}
	\centering \includegraphics[scale=0.15]{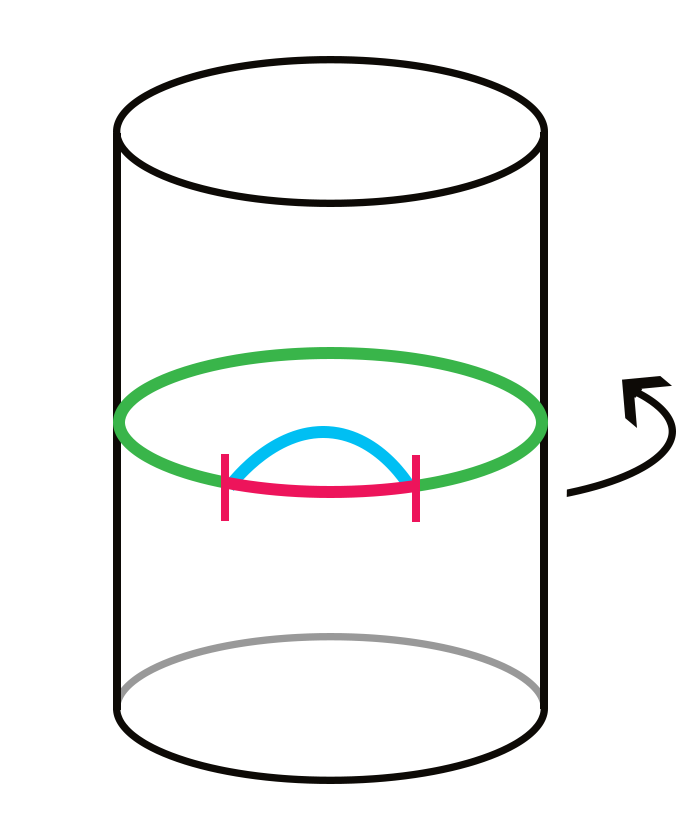}\llap{
		\parbox[b]{3.4cm}{a)\\\rule{0ex}{3.2cm}}} \llap{
		\parbox[][-2.8cm][b]{3.4cm}{$t_E = 0$}} \llap{
		\parbox[][-3.4cm][b]{.6cm}{$x$}} 
	\qquad  
	\includegraphics[scale=0.15]{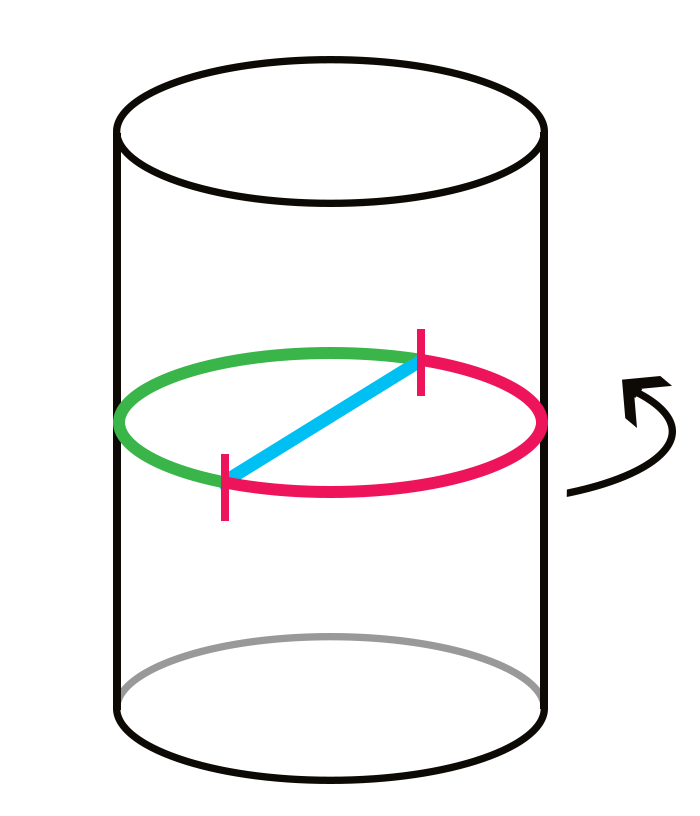}\llap{
		\parbox[b]{3.4cm}{b)\\\rule{0ex}{3.2cm}}} \llap{
		\parbox[][-3.4cm][b]{.5cm}{$x$}} \qquad 
	\includegraphics[scale=0.16]{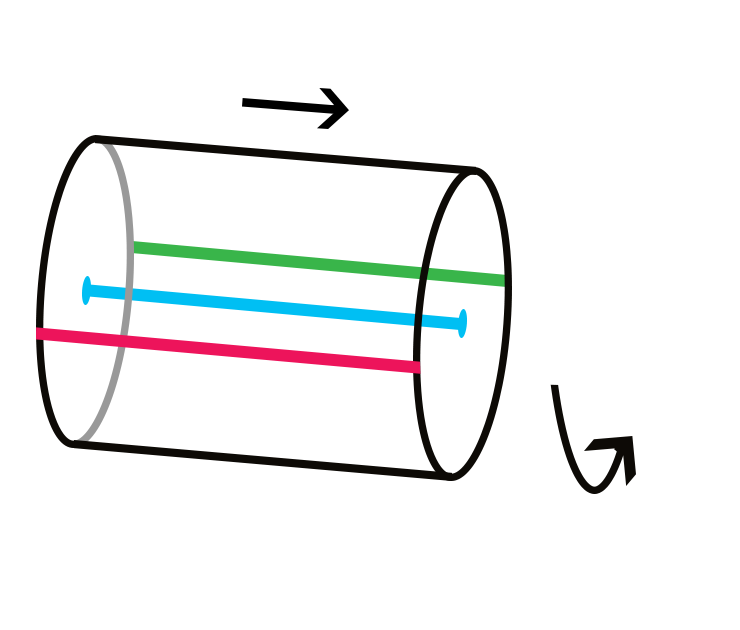}\llap{
		\parbox[b]{3.4cm}{c)\\\rule{0ex}{3.2cm}}} \llap{
		\parbox[][-4.6cm][b]{2.1cm}{$x_z$}} \llap{
		\parbox[][-1.7cm][b]{0.6cm}{$\tau$}}
	\caption{The AdS (in a and b) and BTZ string dual (in c) of the 2-dimensional CFT set-ups in Fig \ref{figHolzheypsi}, in Euclidean signature. They are related by conformal mappings in the boundary or coordinate transformations in the bulk. The CFT region $A$ is in red, the complementary region $\bar A$ in green, and the RT surface $r = r_+$ in blue.} 
	\label{figHolzheyholo} 
\end{figure}

The BTZ string spacetime \eqref{BTZstring} is just a different (namely hyperbolic) slicing of 3-dimensional AdS. 
The explicit coordinate transformation to AdS in Poincar\'e coordinates 
\ali{
	ds^2 = \frac{1}{Z^2} \left(dZ^2 - dt^2 + dx^2 \right) 
} 
is given e.g.~in the appendix of \cite{Asplund:2016koz} (via embedding coordinates). 
It reduces asymptotically to the transformation from Rindler to Minkowski coordinates, $x = e^{x_z} \cosh t_S$ and $t = e^{x_z} \sinh t_S$.  
In BTZ slicing, the constant time slice of AdS$_3$ is presented in Fig \ref{figBTZstringcoord}, with the BTZ string metric \eqref{BTZstring} covering one half. 
Asymptotically, this means the half-space of the plane, which we can call region $A$ as in Fig \ref{figHolzheypsi}b,   
is described by the full range $x_z=-\infty..\infty$ of the BTZ string coordinate. This is pictured in Fig \ref{figHolzheyholo}b and c. Indeed, each of the conformal transformations depicted in Fig \ref{figHolzheypsi} that are needed to map an interval $A$ in a 2-dimensional CFT to the full system size at an auxiliary temperature $T$ is dual to an AdS bulk coordinate transformation. 
This is illustrated in Fig \ref{figHolzheyholo}, which provides the holographic dual interpretation of Fig \ref{figHolzheypsi}, in Euclidean signature. The explicit bulk coordinate transformation from Fig \ref{figHolzheyholo}a to \ref{figHolzheyholo}b, dual to conformally mapping the interval to the half-line, can also be found in \cite{Casini:2011kv}\footnote{
	While we have restricted to the AdS$_3$/CFT$_2$ case for concreteness, the discussion in \cite{Casini:2011kv} holds for any dimension. 
}.

Having arrived at Fig \ref{figHolzheyholo}c, the entanglement $S_A$ for an interval $A$ is equal to the thermal entropy $S_{thermal}$ given by the Bekenstein-Hawking formula \eqref{BHentropy} for the BTZ string geometry with Hawking temperature $T$. That is, $S_A$ is calculated holographically by the area of the BTZ horizon. Under the mappings from Fig \ref{figHolzheyholo}c back to the coordinates of Fig \ref{figHolzheyholo}a, the horizon is mapped to the RT minimal surface, and the RT prescription is proven by having found the bulk transformations dual to the conformal transformations of Holzhey et.~al.~\cite{Holzhey:1994we} combined with the holographic prescription \eqref{BHentropy} for calculating a thermal entropy. 
The length of the RT geodesic is divergent because it extends to the boundary, reflecting as it should the UV divergence of $S_A$. This is consistent with the area of the \emph{planar} BTZ \emph{string} geometry being divergent, as opposed to the finite area of the BTZ black hole (obtained from a quotient of AdS).

From \eqref{Sthermalreplica}, it is clear that the above sketched Casini Huerta Myers proof of the RT conjecture forms a special, U(1) symmetric case
of the bulk replica proof of section \ref{sectionLM}, with the RT surface provided by the U(1) fixed point or more generally the $\mathbb Z_n$ fixed point. 
By continuity, the RT surface ending on the $\mathbb Z_n$ fixed points $\p A$ in the boundary consists of the set of $\mathbb Z_n$ fixed points in the bulk.

\begin{figure}
	\centering \includegraphics[scale=0.16]{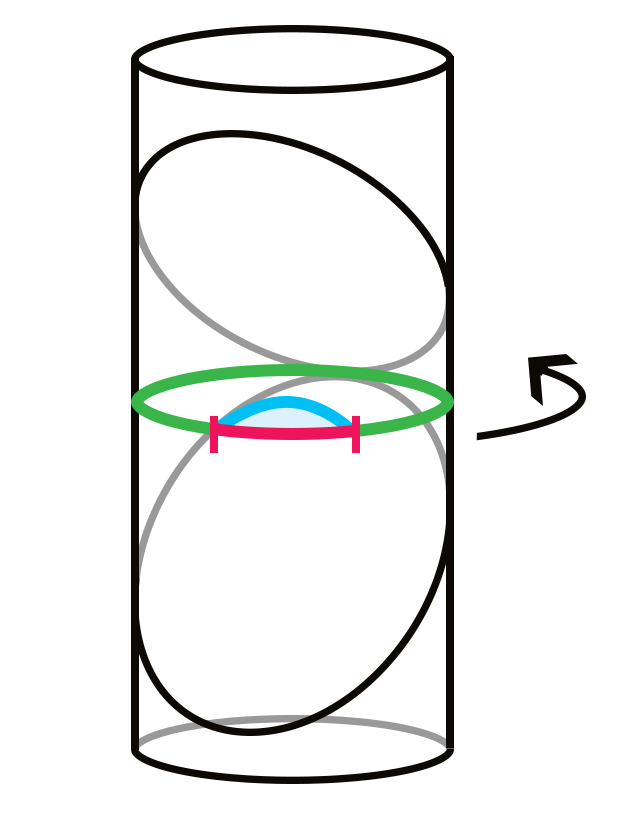}\llap{
		\parbox[b]{3.4cm}{a)\\\rule{0ex}{3.2cm}}} \llap{
		\parbox[][-3.1cm][b]{3cm}{$t=0$}} \llap{
		\parbox[][-3.85cm][b]{.7cm}{$x$}} \qquad \quad 
	\includegraphics[scale=0.16]{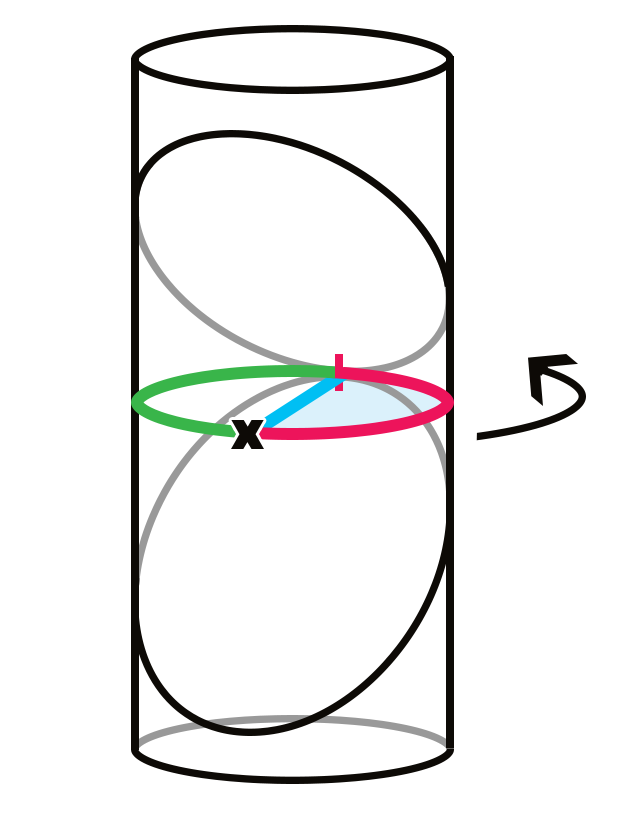}\llap{
		\parbox[b]{3.4cm}{b)\\\rule{0ex}{3.2cm}}} \llap{
		\parbox[][-3.8cm][b]{.65cm}{$x$}}  \qquad \quad
	\includegraphics[scale=0.16]{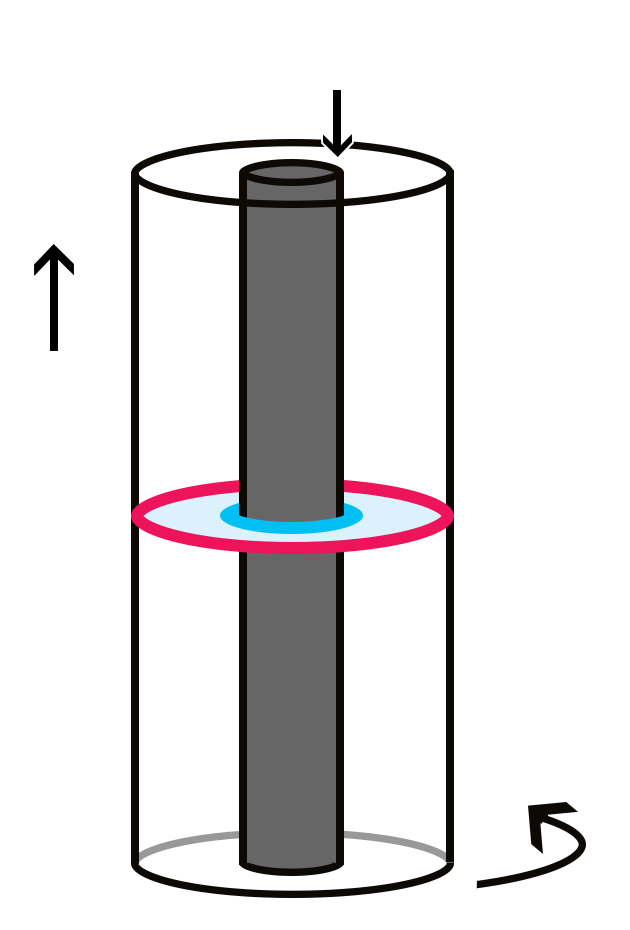}\llap{
		\parbox[b]{3.4cm}{c)\\\rule{0ex}{3.2cm}}} \llap{
		\parbox[][-5.9cm][b]{2.7cm}{$t_S$}} \llap{
		\parbox[][-7.1cm][b]{1.5cm}{$r=r_+$}} \llap{
		\parbox[][-1.1cm][b]{0.8cm}{$x_z$}}
	\caption{The AdS (in a and b) and BTZ string dual (in c) of the 2-dimensional CFT set-ups in Fig \ref{figHolzheypsi}, in Lorentzian signature. The AdS-Poincar\'e coordinates $(t,x)$ cover the region of the cylinder inside the Poincar\'e horizon.  
		The entanglement wedge (light blue shaded) of $A$ (red) maps to the  outside-horizon region of the BTZ string geometry. 
		In the right-most figure, the spatial coordinate has been compactified for reasons of presentation, because in this way it is  
		most clear visually that the Bekenstein-Hawking entropy inspiration behind the RT prescription becomes explicit in the U(1) case. However, in this picture, the coordinate $x_z$ should still be considered to run from $-\infty$ to $\infty$, covering the cylinder infinitely many times, and thus the horizon area in Fig c is still divergent.} 
	\label{figHolzheyholoLor} 
\end{figure}

For each of the Euclidean bulk pictures in Fig \ref{figHolzheyholo}, Fig \ref{figHolzheyholoLor} shows the corresponding Lorentzian one. 
In Lorentzian signature, the Poincar\'e coordinates $(Z,t,x)$ cover the 
region of AdS within the AdS-Poincar\'e horizon. This horizon intersects the AdS boundary at the null boundaries of the Minkowski spacetime Penrose diagram of Fig 6a, with BTZ string coordinates describing the bulk dual of each of the Rindler wedges. 

\section{Gravitational EOM from CFT entanglement} \label{sectemergentgrav}

The RT formula gives an intriguing geometric interpretation to CFT entanglement. Now the bulk geometry has to satisfy Einstein's equations and we present in this section the basic ideas behind entanglement interpretations of the gravitational dynamics. At first we focus on holography, in the last subsection we comment on similar ideas in non-holographic contexts.

\subsection{Holographic emergent gravity}  \label{subsectholograv}

\subsubsection{Bulk metric equation of motion}

We can consider the first law of black hole thermodynamics \cite{Bardeen:1973gs} for the BTZ string black hole of Fig \ref{figHolzheyholoLor}c, 
\ali{
	\delta M = T \delta S_{thermal}, \label{bhfirstlaw} 
} 
with $T = 1/(2\pi)$ (corresponding to the choice of $\kappa = 1$ in   \eqref{BWtheorem}).   
Equation \eqref{bhfirstlaw} expresses how a small change in mass of the black hole is related to its change in entropy. As it relates two solutions of the Einstein equations (one with mass $M$ and one with mass $M + \delta M$), the first law is an on-shell expression that is equivalent to the linearized Einstein equations. This is made precise in the Iyer-Wald formalism \cite{Wald:1993nt,Iyer:1994ys}. 
In essence, that formalism writes the black hole first law \eqref{bhfirstlaw} as a Stokes equation $\delta E_\infty - \delta E_h = 0$, with $E_\infty$ and $E_h$ the energy at infinity and the horizon. However, it is valid far more broadly than the case above, 
allowing to write a gravitational first law expressing energy conservation in a chosen region of a solution of the equations of motion, for any gravitational theory with a diffeomorphism invariant Lagrangian. For example, the Iyer-Wald formalism can be applied directly to the entanglement wedge region in Fig \ref{figHolzheyholoLor}a. In fact, we will shortly interpret the black hole first law \eqref{bhfirstlaw} as the gravitational first law for the entanglement wedge of $A$. We are focusing here for concreteness on the most intuitive U(1) symmetric case, where the entanglement wedge maps to a black hole outside-horizon region.

One can now ask what the dual CFT interpretation of the gravitational law  \eqref{bhfirstlaw} is. 
By the Bekenstein-Hawking formula \eqref{BHentropy} and the mappings in Fig \ref{figHolzheyholoLor}, i.e.~by the RT formula of AdS/CFT, the right hand side of \eqref{bhfirstlaw} is equal to $S_A/(2\pi)$, the vacuum entanglement entropy of region $A$. On the left hand side, the mass $M$ is obtained from integrating the Brown-York stress tensor, which is identified in AdS/CFT with the expectation value of the CFT stress tensor \cite{Balasubramanian:1999re}, so that $\delta M = \delta \vev{H_{t_S}}$ in terms of the boundary 
Rindler Hamiltonian.  
Because of the Bisognano Wichmann result \eqref{BWtheorem} (with $\tau$ the  Euclidean Rindler time $\tau = -i t_S$),  
the Rindler Hamiltonian is  equal (up to a normalization constant) to $1/(2\pi)$ times the `modular Hamiltonian' $H_{mod,A}$, which is an operator defined by writing the positive semi-definite reduced density matrix $\rho_A$ of the vacuum CFT state in the form 
\ali{
	\rho_A = \frac{e^{-H_{mod,A}}}{\tr e^{-H_{mod,A}}}. 
}

Having used the AdS/CFT dictionary on each side of the black hole first law \eqref{bhfirstlaw}, it now reads, in terms of CFT entanglement concepts:  
\ali{
	\delta \vev{H_{mod,A}} = \delta S_A.  \label{entfirstlaw}
} 
This is a relation known as the `first law of entanglement' \cite{Bianchi:2012br,Wong:2013gua,vR-Lashkari:2013koa}, 
for its close resemblance, and in this case connection, to the first law of thermodynamics. 
The conclusion is that a gravitational first law in the bulk, namely 
the one associated with the entanglement wedge region of $A$, corresponds to an entanglement first law in the CFT. It follows that from the CFT perspective, the entanglement first law \eqref{entfirstlaw} imposes linearized Einstein equations in the dual bulk theory. 
In this sense gravity can be said to emerge from entanglement in AdS/CFT   \cite{VanRaamsdonk:2010pw,vR-Lashkari:2013koa,vR-Faulkner:2013ica}. 

Bulk matter can be included in a semi-classical treatment of the bulk theory, leading to an additional matter energy contribution to the first law \eqref{bhfirstlaw}, $\delta M = T \delta S + \delta E_m$. The gravitational first law is then still consistent with the first law of entanglement \emph{if} the RT prescription is corrected to include a bulk matter entropy contribution, which is indeed the proposal of the FLM \cite{FLM-Faulkner:2013ana} and QES \cite{Engelhardt:2014gca} prescriptions.

\subsubsection{Bulk matter equation of motion}

We discussed how equations of motion for the geometry in a gravitational theory can be thought of as having an entropic origin. One can also ask about the equations of motion for bulk \emph{matter} degrees of freedom, 
and to what extent these can be described 
using statements about boundary entanglement. 
Considering for concreteness a free bulk scalar field $\phi$, we can consider how it behaves under the action of the total CFT modular Hamiltonian, defined for a given CFT region $A$ as the difference $H_{mod,A}^{tot} \equiv H_{mod,A} - H_{mod,\bar A}$. It is an operator that can be written as a linear combination of the CFT conformal generators $L_0$, $L_1$ and $L_{-1}$ \cite{deBoer:2016pqk}. One obtains what is known as the JLMS result  \cite{Jafferis:2015del, Kabat:2018smf}
\ali{
	[H_{mod,A}^{tot}, \phi] = \mathcal L_\xi \phi   \label{JLMS}
	}
with $\mathcal L_\xi$ the Lie derivative in the direction of $\xi$, the AdS Killing vector acting in the entanglement wedge of $A$ and vanishing on the RT surface. 
One can then argue \cite{Callebaut:2022mqw} that the bulk equation of motion for $\phi$, written in the form $EOM = 0$, should also satisfy 
\ali{
	[H_{mod,A}^{tot}, EOM] = \mathcal L_\xi (EOM).  \label{Hmodcondition}
}
This gives a consistency condition between a modular Hamiltonian condition in the CFT and the bulk matter equation of motion, strongly constraining the latter. For electromagnetically or gravitationally interacting bulk scalar fields, the right hand side of the JLMS condition \eqref{JLMS} gets corrected, reflecting the non-locality of such bulk objects. In particular, the bulk equation of motion for a gravitationally interacting scalar field in three dimensions satisfies \eqref{Hmodcondition} with $\xi$ equal to $\xi_{Ban}$, the asymptotic Killing vector of the Ba\~{n}ados geometry \cite{Banados:1998gg}.

\subsection{Non-holographic emergent gravity} \label{subsectnonholo}

The idea of a connection between Einstein's equations and entropy laws goes back to the seminal work of Jacobson \cite{Jacobson:1995ab}. A modern entanglement entropy version of it is discussed in \cite{Jacobson:2015hqa} and describes how a given CFT (in dimensions $d>2$) can be coupled to a theory of gravity in the same number of dimensions by imposing the `maximal vacuum entanglement condition'. This statement, 
similarly to the statements on holographic emergent gravity from entanglement in section \ref{subsectholograv}, can be traced back to a reinterpretation of a gravitational first law in terms of entanglement \cite{Jacobson:2015hqa,Bueno:2016gnv}. A 2-dimensional version of Jacobson's emergent gravity from maximal vacuum entanglement has been discussed in \cite{Callebaut:2018xfu}, see also \cite{Pedraza:2021cvx,Pedraza:2021ssc}, 
and a closely related account on entropic emergence of Jackiw-Teitelboim gravity for a CFT with boundary was presented in \cite{CallebautVerlinde:2018nlq}.

\bibliographystyle{JHEP} 
\bibliography{referencesReview} 

\begin{acknowledgement}
	I would like to thank my collaborators, and more broadly my friends and colleagues, in the fascinating world of entanglement and holography. 
	Special thanks to Vincent Callebaut for the figures. Funded, in part, by the Deutsche Forschungsgemeinschaft (DFG, German Research Foundation) – Projektnummer 277101999 – TRR 183 (project A03 and B01). \\
	
	This is a preprint of the following chapter: N. Callebaut, Entanglement in Conformal Field Theory and Holography, published in Gravity, Cosmology, and Astrophysics: A Journey of Exploration and Discovery with Female Pioneers, edited by B. Hartmann and J. Kunz, 2023, Springer, reproduced with permission of Springer Nature Switzerland AG. The final authenticated version is available online at: \url{http://dx.doi.org/10.1007/978-3-031-42096-2_10}.   
\end{acknowledgement}

\backmatter


\end{document}